\documentclass{jkas}
\def\year{2023} 
\def\month{Feb} 
\def\volume{00} 
\def\issue{0} 
\def\beginpage{1} 
\def\endpage{99} 
\def\received{---} 
\def\accepted{---} 
\date{Received \received; accepted \accepted}

\usepackage{amssymb,amsmath,color,newtxtext,newtxmath,graphicx,pdflscape}

\title{A Deep Optical Survey of Young Stars in the Carina Nebula. I. \\
 -- {\it UBVRI} Photometric Data and Fundamental Parameters}

\author[1]{Hyeonoh Hur}
\author[2,3,4]{Beomdu Lim}
\author[4]{Moo-Young Chun}
\affil[1]{The Young Astronauts Korea, Office Number 407, 22 Teheran-ro 7-gil, Gangnam-gu, Seoul, Korea}
\affil[2]{Department of Earth Science Education, Kongju National University, 56 Gongjudaehak-ro, Gongju-si, Chungcheongnam-do 32588, Republic of Korea \email{blim@kongju.ac.kr}}
\affil[3]{Earth Environment Research Center, Kongju National University, 56 Gongjudaehak-ro, Gongju-si, Chungcheongnam-do 32588, Republic of Korea }
\affil[4]{Korea Astronomy and Space Science Institute, 776 Daedeok-daero, Yuseong-gu, Daejeon 34055, Republic of Korea}

\def\corrauthor{Beomdu Lim}
\def\runningauthor{Hur et al.}
\def\runningtitle{An Optical View of the Carina Nebula}
\def\keywords{open clusters and associations: NGC 3372, nebulae, ISM: HII regions }

\def\abstracttext{
We present the deep homogeneous $UBVRI$ photometric data of 
135,071 stars down to $V\sim23$ mag and $I\sim22$ mag toward 
the Carina Nebula. These stars are cross-matched with those 
from the previous surveys in the X-ray, near-infrared, and 
mid-infrared wavelengths as well as the Gaia Early Data 
Release 3 (EDR3). This master catalog allows us to select reliable 
members and determine the fundamental parameters distance, size, stellar density 
of stellar clusters in this star-forming region. We revisit the 
reddening toward the nebula using the optical and the 
near-infrared colors of early-type stars. The foreground reddening [$E(B-V)_{fg}$] is determined to be $0.35\pm0.02$, and it 
seems to follow the standard reddening law. On the other 
hand, the total-to-selective extinction ratio of the intracluster 
medium ($R_{V,cl}$) decreases from the central region 
(Trumpler 14 and 16, $R_{V,cl}\sim4.5$) to the northern region 
(Trumpler 15, $R_{V,cl}\sim3.4$). It implies that the central 
region is more dusty than the northern region. We find that 
the distance modulus of the Carina Nebula to be $11.9\pm0.3$ mag 
($d=2.4\pm0.35$ kpc) using a zero-age main-sequence fitting method, 
which is in good agreement with that derived from the Gaia 
EDR3 parallaxes. We also present the catalog of 3,331 pre-main-sequence (PMS) 
members and 14,974 PMS candidates down to $V \sim 22$ 
mag based on spectrophotometric properties of young stars 
at infrared, optical, and X-ray wavelengths. From the 
spatial distribution of PMS members and PMS candidates, we confirm 
that the member selection is very reliable down to faint stars.
Our data will have a legacy value for follow-up studies 
with different scientific purposes.}

\begin{document}
\jkashead 

\section{Introduction}
Massive stars have strong influence on the formation and 
early evolution of low-mass stars via producing enormous 
amount of mass outflow, strong stellar winds, and 
far-ultraviolet photons. The Carina Nebula (NGC 3372, 
R.A. $= 10:45:02$, decl. = $-59:41:60$, J2000) 
is one of ideal laboratories to study the influence 
of massive stars on low-mass star formation. There are 
a large number of massive stars over the nebula, and 
most of them are found in several rich clusters \citep{sb08,so14}. 
Trumpler 14 (Tr 14) and Trumpler 16 (Tr 16) are in the 
brightest part of the Carina Nebula. There is a luminous 
blue variable (LBV) star, $\eta$ Carinae. This massive 
star is one of the most interesting massive stars in 
the Galaxy because of its unusual historical variability 
\citep{sm11} and surrounding lobe-shape ejecta. Trumpler 15 
(Tr 15) is also a rich cluster and is believed to be 
physically associated with Tr 14 and Tr 16 \citep{fe11}. 
Collinder 232 (Cr 232), close 
to the Tr 14, is a less dense stellar group in the 
optical wavelength \citep{hur12} (hereafter HSB12), it contains an embedded 
group \citep{po11a}. 

There have been large surveys at various wavelengths 
for the study of the stellar content in the Carina 
Nebula. The {\it Chandra} Carina Complex Project 
(CCCP, \citealt{to11}) published a catalog of X-ray 
sources within an area larger than a square degree. 
This catalog includes source classifications based 
mainly on median X-ray energy, variability in X-ray, 
and infrared excess \citep{br11a}. Its X-ray detection 
limit reaches down to $\sim 1$ M$_{{\odot}}$ pre-main-sequence 
(PMS) stars in the nebula \citep{pr11}. At near-infrared 
(NIR) wavelengths, $JHK_S$ photometric survey data obtained from 
the Visible and Infrared Survey Telescope for Astronomy 
(VISTA) were published by \citet{pr14}, which covers 6.7 
square degrees down to J$\sim$ 21 mag. The Vela-Carina survey at the mid-infrared (MIR) bands 
with the {\it Spitzer}/IRAC (the Infrared Array Camera) 
\citep{ma07} are very useful to select PMS stars with 
circumstellar disks in the Carina Nebula. The Galactic 
O-Star Spectroscopic Survey (GOSSS) \citep{so14} provided 
homogeneous spectroscopic classifications for the known 
O-type stars.

Such extensive multi-wavelength surveys provide several 
decisive selection criteria for young PMS stars, but 
there is no deep optical photometric data complete down 
to low-mass stars. There are several previous 
optical CCD photometric studies (\citealt{mj93,va96,ca02,ca04,ta03}; HSB12) using 1m-class telescopes. 
In such survey, only a small field of view 
(about $20^{\prime}\times20^{\prime}$) was covered. 
Furthermore, systematic differences among these previous 
CCD photometric data were reported by our previous study 
(HSB12). A deep and homogeneous optical photometric survey 
of this star-forming complex is required to study low-mass 
members in multi-wavelength bands.

In this paper, we present deep homogeneous optical 
photometric data of a larger than a $36.0^{\prime}\times36.6^{\prime}$ 
area centered on the brightest part of the Carina Nebula. 
Our data are very useful to select genuine members and 
study their physical parameters when combined with the 
previous sets of multi-wavelength data. In Section 2, we 
describe how the data were obtained and combined into 
a homogeneous data set. Cross identifications between 
our new optical data and the data from previous surveys 
are described in Section 3. In Sections 4 and 5, the 
reddening, its spatial variation, and distance are 
described. Membership selection criteria for low-mass PMS stars from 
color-magnitude diagrams (CMDs) and two-color diagrams 
(TCDs) are addressed 
in Section 6. The spatial distribution of the selected 
members and member candidates are presented and discussed 
in Section 7. Finally, our results are summarized in Section 8.

\begin{table*}
\centering
\caption{The Extinction Coefficients and Photometric Zero-points for the CTIO 4m Data}
\label{ctio4m_extc}
  \begin{tabular}{ccccccc}
  \hline
  \hline
Filter & $k_1$             & Color &  $\zeta_{chip1}$   & $\zeta_{chip2}$     & $\zeta_{chip3}$    & $\zeta_{chip4}$    \\
       &                   &       &  $\zeta_{chip5}$   & $\zeta_{chip6}$     & $\zeta_{chip7}$    & $\zeta_{chip8}$    \\
  \hline
$V$    & 0.137 $\pm$ 0.009 & $V-I$ & 25.638 $\pm$ 0.010 & 25.618 $ \pm$ 0.010 & 25.620 $\pm$ 0.009 & 25.639 $\pm$ 0.010 \\
       &                   &       & 25.653 $\pm$ 0.009 & 25.642 $ \pm$ 0.009 & 25.626 $\pm$ 0.008 & 25.645 $\pm$ 0.008 \\
$R$    & 0.083 $\pm$ 0.011 & $R-I$ & 25.765 $\pm$ 0.010 & 25.742 $ \pm$ 0.010 & 25.736 $\pm$ 0.012 & 25.778 $\pm$ 0.010 \\
       &                   &       & 25.774 $\pm$ 0.012 & 25.738 $ \pm$ 0.009 & 25.731 $\pm$ 0.010 & 25.776 $\pm$ 0.012 \\
$I$    & 0.060 $\pm$ 0.008 & $V-I$ & 25.080 $\pm$ 0.011 & 25.042 $ \pm$ 0.012 & 25.048 $\pm$ 0.012 & 25.086 $\pm$ 0.012 \\
       &                   &       & 25.092 $\pm$ 0.011 & 25.074 $ \pm$ 0.012 & 25.040 $\pm$ 0.012 & 25.088 $\pm$ 0.013 \\
$I$    & 0.060 $\pm$ 0.008 & $R-I$ & 25.081 $\pm$ 0.011 & 25.044 $ \pm$ 0.012 & 25.050 $\pm$ 0.012 & 25.087 $\pm$ 0.012 \\
       &                   &       & 25.093 $\pm$ 0.011 & 25.073 $ \pm$ 0.012 & 25.042 $\pm$ 0.012 & 25.089 $\pm$ 0.012 \\
  \hline
 \end{tabular}
\end{table*}

\section{Observations}
Tr 14 and Tr 16 had been observed by using the 1m telescope 
at Siding Spring Observatory (SSO) and a thinned SITe 2K 
CCD camera (see HSB12). Here, we describe two additional 
observations at SSO and Cerro Tololo Inter-American 
Observatory (CTIO). 

\begin{figure}
\includegraphics[width=\columnwidth]{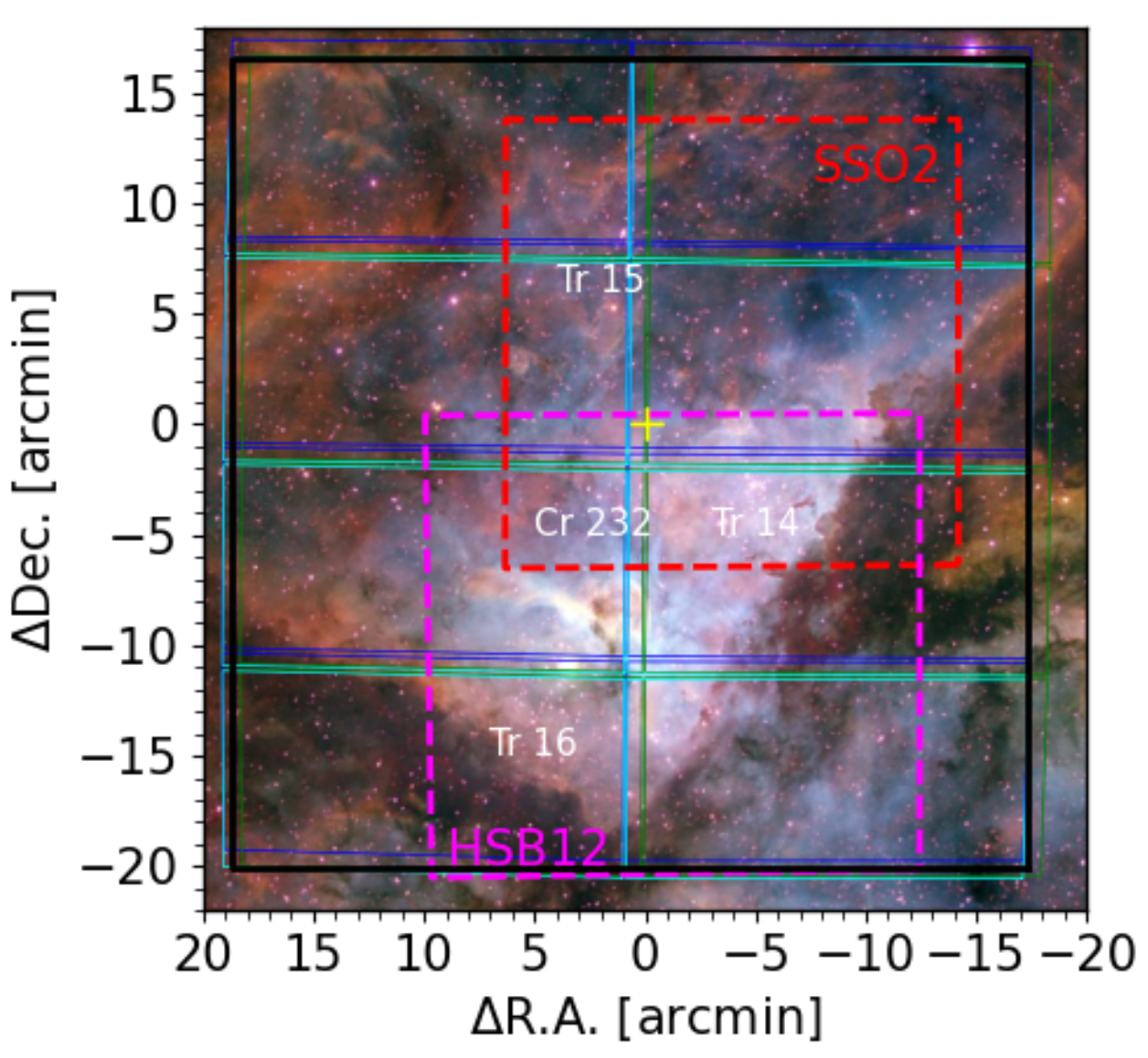}
\caption{Optical image of the Carina Nebula (from: https://noirlab.edu/public/images/noao0136a, 
credit: CTIO/NOIRLab/NSF/AURA/N. Smith (University of Arizona)).
The thin solid lines outline the FOVs of the CTIO 4m MOSAIC II CCD chips 
(blue -- dithering pattern \#1, green -- dithering pattern \#2, and cyan -- dithering pattern \#3), and the thick solid line (black) shows 
the gap-free FOV. The thick dashed lines are the FOVs of the SSO data 
(bottom -- HSB12, top -- SSO2). The coordinates of stars are relative 
to R.A.$_{J2000}=$10:44:36.00 and decl.$_{J2000}= $-59:30:00.0 (yellow cross).
}
\label{chart}
\end{figure}

\subsection{SSO 1m Observation}
The imaging observations of Tr 15 were performed 
on 1997 February 28 (hereafter SSO2) using the same 
instrumental setup as that used by HSB12. The exposure 
times were 2$\times$300s and 30s in $U$, 2$\times$100s 
and 10s in $B$, and 2$\times$60s and 5s in $V$ and $I$, 
respectively. The average seeing was $\sim1.6^{\prime\prime}$. 
We performed the point-spread-function 
(PSF) fitting photometry using the {\tt daophot} package in {\tt iraf}. An aperture correction was applied to 
adjust the magnitudes obtained from simple aperture 
photometry with a radius of 7$^{\prime\prime}$. The instrumental 
magnitudes were transformed to the Landolt's standard 
system according to the transformation relations 
obtained by \citet{su98}.

\subsection{CTIO 4m Observation \label{ctio4m} }
Deep imaging observations were performed at CTIO 
using the 8k$\times$8k MOSAIC II CCD camera attached 
to the 4m Blanco telescope and $VRI$ filters on 2008 
May 6. We took two sets of exposure times to cover 
both bright and faint stars. The exposure times for 
faint stars were set to 150s, 150s, and 75s in 
$V$, $R$, and $I$ bands, respectively, while we 
took some frames to observe bright stars with exposure 
time of 3s in all bands. In order to fill in the gaps 
between the chips of the MOSAIC II CCD camera, three 
frames for each setup were taken by our own dithering 
pattern. The gap-free FOV was $-17.36\leq\Delta 
\mathrm{R.A.}[^{\prime}]\leq 18.65$ and $-20.03\leq\Delta {\mathrm{decl.}} [^{\prime}]\leq 16.59$ ($36.01^{\prime} \times 36.62^{\prime}$) relative 
to the equatorial coordinate (R.A., decl.)$_{\text{J2000}} = $(10:44:36.00, -59:30:00.0). 
The FOVs of the observations are shown in Figure~\ref{chart}. 
The average seeing was better than $\sim 1.0^{\prime\prime}$. We also 
observed a large number of the standard stars in the 
Landolt's regions \citep{la92,st00} using the same filter 
set on the same night. Additional observations of the standard regions 
were performed to derive the reliable standard transformation 
relations in $V$ and $I$ bands on 2009 March 28 and 29 (see also 
\citealt{li13,hur15}).

\begin{figure*}
\includegraphics[width=2.1\columnwidth]{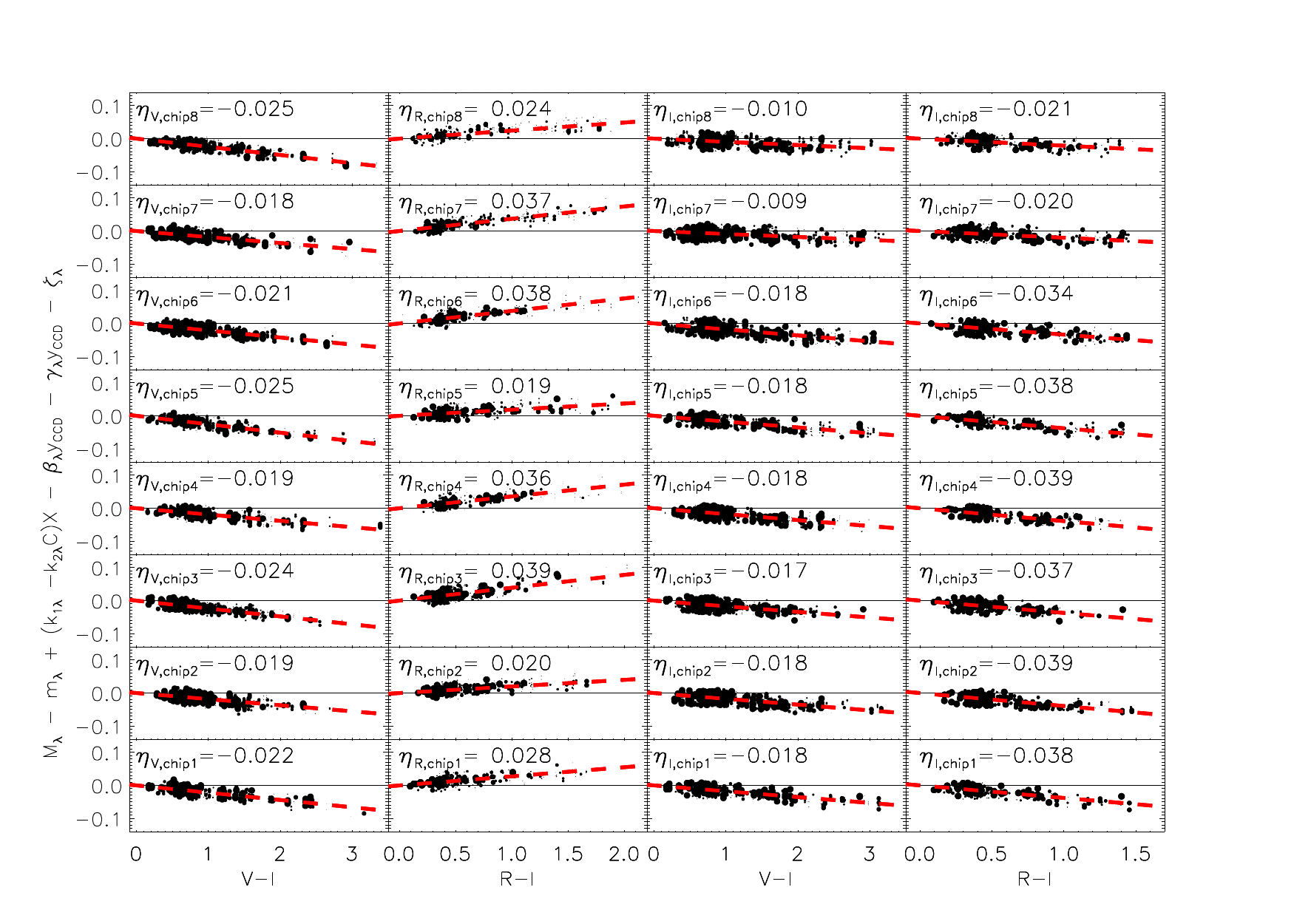}
\caption{The coefficients of the MOSAIC II CCD camera for the transformation to the Landolt system.
The size of dots is based on the reliability of the data.}
\label{col_coeff}
\end{figure*}

\begin{figure*}
\includegraphics[angle=90,width=1.9\columnwidth]{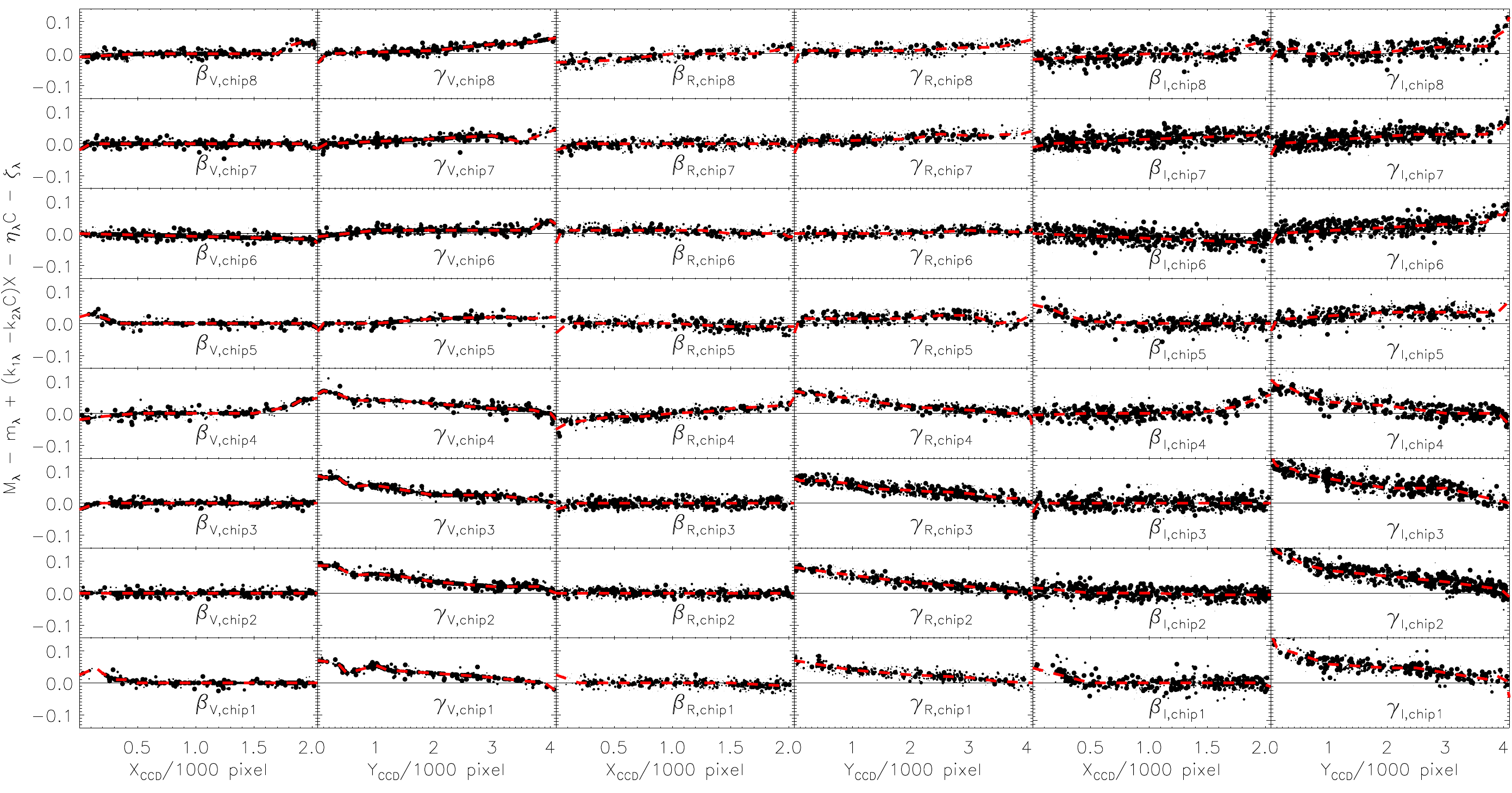}
\centering
\caption{
The spatial variation of the photometric zero-points of the MOSAIC II CCD camera.
Each panel shows zero-point dependence of $V$ (left two columns), $R$ (middle two columns) and $I$ (right two columns) magnitudes on X and Y coordinates.
The size of dots is based on the reliability of the data.}
\label{spatial_coeff}
\end{figure*}

The observed images were pre-processed to remove the 
instrumental signatures by conducting bias subtraction, 
flat fielding, and cross-talk correction using the \textsc{mscred} 
package in \textsc{iraf}\footnote{\: Image Reduction and 
Analysis Facility is developed and distributed by the 
National Optical Astronomy Observatories which is operated
by the Association of Universities for Research in Astronomy 
under cooperative agreement with the National Science 
Foundation.}. We performed simple aperture photometry for 
the observed standard stars using an aperture radius of 
$5^{\prime\prime}$. The transformation relations 
to the standard system for the MOSAIC II camera have been 
derived by \citet{li13} and \citet{hur15}. Although the 
previous relations also reproduce the Landolt's standard 
system well, we slightly modified the spatial variation 
coefficients to correct a non-linear spatial variation 
of photometric zero points at the edge of each chip 
to minimize chip-to-chip differences 
in magnitudes of the same stars. The general transformation 
relation we modified \citep{su08} is:

\begin{equation}
\label{std_eq}
  \begin{split}
   M_{\lambda} = m_{\lambda} - (k_{1\lambda} -k_{2\lambda}C)X + \eta_{\lambda}C \\
   + \alpha_{\lambda}\hat{UT} + \beta_{\lambda}X_{CCD} + \gamma_{\lambda}Y_{CCD} +\zeta_{\lambda} 
  \end{split}
\end{equation}
\noindent where $M_{\lambda}$, $m_{\lambda}$, $k_{1\lambda}$, 
$k_{2\lambda}$, $C$, $X$, $\eta_{\lambda}$, $\alpha_{\lambda}$, 
$\hat{UT}$, $\beta_{\lambda}$, $\gamma_{\lambda}$, and $\zeta_{\lambda}$ 
represent the standard magnitude, instrumental magnitude after aperture 
correction, the primary extinction coefficient, the secondary extinction coefficient, relevant color index, airmass, transformation coefficient 
to the standard system, time-variation coefficient, time difference 
relative to the midnight, spatial variation coefficients along the 
X and Y coordinates of the MOSAIC II chips in units of 1000 pixel, 
and photometric zero-point, respectively. The secondary extinction 
coefficient is, in general, close to zero for $V$, $R$, and $I$ passbands, 
and so we adopted $k_{2\lambda} = 0$. The photometric zero points 
during the observing nights were very stable ($\alpha_{\lambda}\hat{UT}=0$). 
The transformation coefficients (dashed lines in Figure~\ref{col_coeff}) were determined by using a least square 
fitting method. The spatial variation of photometric zero-points 
is associated with the variation of pixel scales across the wide 
FoV. We probed the spatial variation of photometric zero-points 
using both standard stars and stars in the Carina Nebula. As a result, 
it appears as a non-linear pattern in each band as shown in Figure~\ref{spatial_coeff}. We traced the such a non-linear pattern 
by adopting an empirical curve passing through roughly middle of 
the data points at given positions. The other coefficients are 
summarized in Table~\ref{ctio4m_extc}.

For stars in the Carina Nebula, we performed the PSF photometry 
using the \textsc{daophot} package in \textsc{iraf}. In order to 
minimize inclusion of spurious sources, probable spurious sources 
were rejected based on their photometric error, $\chi$, sharpness 
values, and their surface profiles. Subsequently, all faint 
sources (photometric error larger than 0.1 mag, $\epsilon>0.1$ mag) and extended sources (sharpness $>0.5$) 
in the original images were visually checked. 
We also generated median images smoothed by a $10\times10$ square 
pixel. The original images were then subtracted by the median images. 
We searched for several definite point sources embedded in bright 
nebulae or the extended PSF wing of bright stars in these 
median-subtracted images. The magnitudes of these stars were 
obtained in the same manner as above. 

Finally, an aperture correction was conducted to derive the 
equivalent magnitude for the same aperture size 
as that of the standard star photometry ($r = 5^{\prime\prime}$). 
The aperture corrected magnitudes of stars were transformed to 
the Landolt's standard system \citep{la92,st00} using our 
transformation relations. There were a large number of faint 
sources (39,779) detected only in $I$. In 
order to obtain standard $I$ magnitudes of these faint stars, we 
assumed a $R-I$ color of 1.5 as their minimum $R-I$ color. 
In this case, the systematic errors in the standard $I$ magnitude 
is thus less than 0.02 mag ($\sim 0.5\times \eta_{I}$) 
at $R-I = 2$. 

The CCD coordinates of the combined optical photometric 
data were transformed to the equatorial coordinates 
using the Two Micron All Sky Survey (2MASS) point-source 
catalog \citep{sk06}. The RMS errors in the coordinate 
transformations are $\sim0.05$ arcsec in both 
$\alpha_{\text{J2000}}$ and $\delta_{\text{J2000}}$ for 
stars with a good signal-to-noise (S/N) ratio. The position 
errors of fainter stars ($I > 20.0$, $R > 21.5$, or $V>23.0$ mag), 
photometric doubles, or stars in the bright portion of 
the nebula could be larger than this. 

\begin{table*}
\caption{The Combined $UBVRI$ Photometric Data }
\label{table_data}
  \begin{tabular}{p{.6cm}p{1.2cm}p{1.47cm}p{.6cm}p{.6cm}p{.82cm}p{.69cm}p{.57cm}p{.56cm}p{.41cm}p{.41cm}p{.54cm}p{.52cm}p{.50cm}p{.47cm}p{.55cm}p{0.55cm}}
  \hline
  \hline
ID & $\alpha_{J2000}$ & $\delta_{J2000}$ & $V$ & $I$ & $U-B$ & $B-V$ & $V-I$ & $R-I$ &
$\epsilon_V$ & $\epsilon_I$ & $\epsilon_{(U-B)}$ & $\epsilon_{(B-V)}$ & $\epsilon_{(V-I)}$ & $\epsilon_{(R-I)}$ &
N(Obs)$^{1}$ & Remark$^{2}$ \\
  \hline
 24977&10:43:15.51&-59:47:07.0&18.404&16.625&      &     &1.779&0.894&0.017&0.008&     &     &0.019&0.018&220022&  \\
 38714&10:43:47.43&-59:31:26.4&13.537&13.007&-0.161&0.337&0.530&0.311&0.006&0.014&0.028&0.009&0.015&0.021&212611&ED\\
 44874&10:44:00.78&-59:17:10.6&22.624&20.325&      &     &2.299&1.126&0.166&0.031&     &     &0.169&0.036&230023&  \\
 49408&10:44:10.83&-59:39:50.8&      &19.031&      &     &     &1.597&     &0.013&     &     &     &0.013&050002&  \\
 61317&10:44:35.92&-59:23:35.6&12.632&12.213&-0.317&0.282&0.419&0.243&0.008&0.013&0.009&0.013&0.015&0.015&433331&E \\
 62149&10:44:37.66&-59:23:07.3&13.231&12.746&-0.190&0.319&0.485&0.272&0.004&0.004&0.007&0.013&0.006&0.011&433333&E \\
 62439&10:44:38.19&-59:24:44.0&14.287&12.716&-0.144&0.950&1.571&0.835&0.012&0.003&0.014&0.008&0.012&0.009&433333&  \\
 62676&10:44:38.65&-59:30:07.7&18.435&16.445&      &0.974&1.990&0.966&0.064&0.035&     &0.070&0.073&0.051&760266&2 \\
 70049&10:44:54.76&-59:44:03.2&20.691&18.115&      &     &2.575&1.245&0.033&0.004&     &     &0.033&0.009&230023& D\\
 71927&10:44:58.78&-59:49:21.1&11.450&11.241&-0.476&0.123&0.209&     &0.006&0.010&0.009&0.008&0.012&     &441440&e \\
 85757&10:45:28.59&-59:26:46.3&      &20.949&      &     &     &     &     &0.027&     &     &     &     &030000&  \\
 91223&10:45:39.66&-59:40:08.1&18.505&15.409&      &     &3.096&1.453&0.007&0.004&     &     &0.008&0.010&760066&3 \\
127443&10:46:51.63&-59:26:44.4&17.219&15.967&      &     &1.252&0.622&0.011&0.013&     &     &0.017&0.019&220022& D\\
  \hline
 \end{tabular}
  \begin{tabular}{lll}
Column (1): ID of stars, Columns (2) and (3): Equatorial coordinates right ascension and declination.\\
Columns (4) and (5): Magnitudes in $V$ and $I$ bands. Columns (6) to (9): Colors.\\
Columns (10) -- (15): Associated photometric errors. Column (16): Number of the detected images all observations.\\
Column (17): Identification of stars. D -- photometric double stars, numbers -- YSO Class 0/I(1), II(2),  III(3) stars. \\
E -- early-type members with $E(B-V)\geq0.35$, e -- foreground early-type stars with $E(B-V)<0.35$ (See Section 4.2).\\
The entire table is published in the electronic edition of the journal.
A portion is shown here for guidance regarding its form and content. \\
 \end{tabular}
\end{table*}

\subsection{The Combined Optical Data}
There are two photometric systems reproducing the Johnson-Cousins 
$UBV(RI)_C$ standard photometric systems. The one is the South African 
Astronomical Observatory (SAAO) standard system \citep{co78,me89,me91,su00}, 
and the other one is Landolt's system \citep{la92}. The former one 
provides the most precise photometric catalog of standard 
stars (e.g., standard stars in E-region), and it is known to be 
very closely tied to the Johnson's $UBV$ system. For this reason, 
we have a preference for observing the SAAO standard stars (HSB12). On 
the other hand, \citet{la92} published a large number of equatorial 
standard stars in a wide color range within small regions. 
Landolt's standard stars are optimized for observations with modern 
CCD cameras. 

\citet{be95} reported that the differences 
of colors between the Landolt and SAAO standard systems show 
systematic curve-like features (see figures 3 and 4 in his paper). 
The size of such systematic differences is non-negligible in some 
color ranges. Furthermore, there is an additional non-linear 
pattern for A0V stars in $U-B$ color differences (see 
also \citealt{su00,LSB09}). 

\begin{figure}
\centering
\includegraphics[width=0.8\columnwidth]{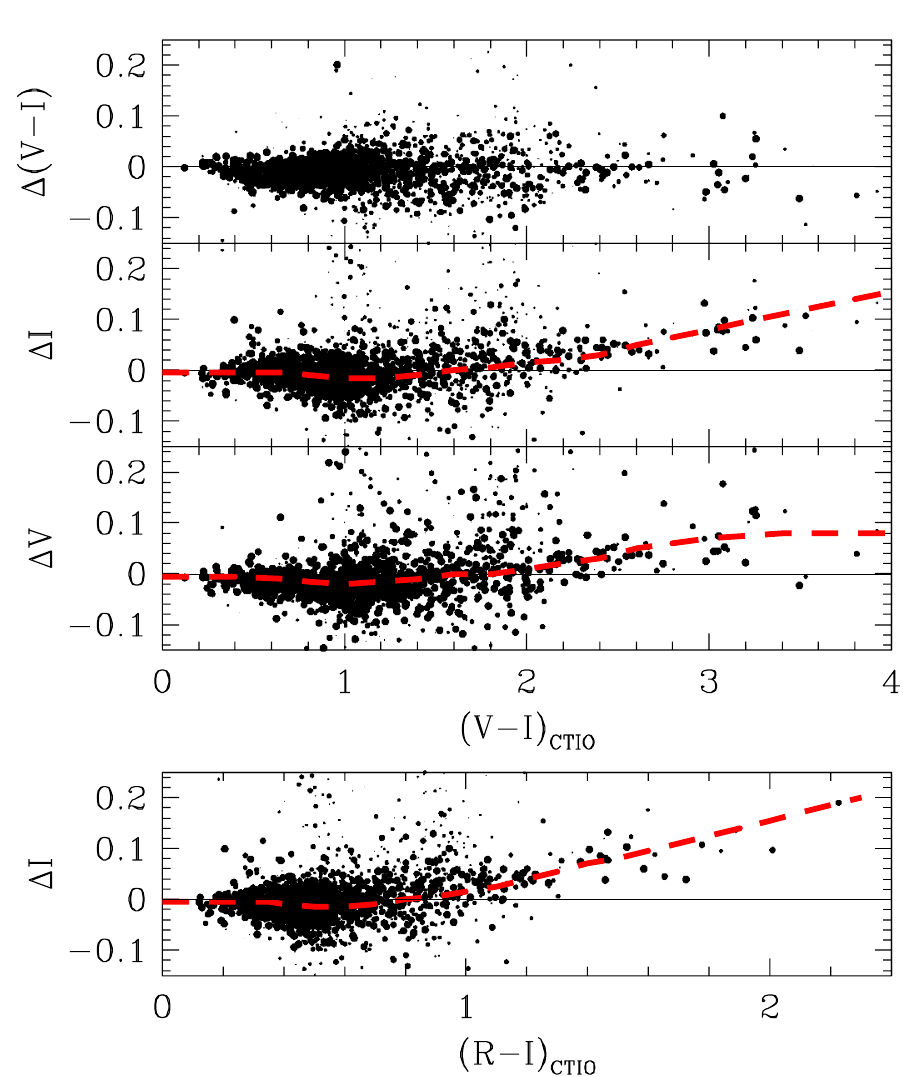}
\caption{
Comparison of photometry -- the CTIO 4m data (the Landolt system) minus the data in HSB12 (the SAAO system).
The larger dots indicate stars with small photometric errors.}
\label{comp_4m_sso}
\end{figure}

We cross-matched the CTIO data (the Landolt standard system) 
with the HSB12 data (the SAAO standard system) for stars with 
$V\leq 17.5$ mag. Figure~\ref{comp_4m_sso} shows the systematic 
differences between the data sets. The $V$ and $I$ magnitudes 
of red stars ($V-I > 2.0$ or $R-I > 1.0$) from the CTIO data 
are systematically fainter than those of stars from SAAO system. 
A similar difference in $V$ is also seen in \citet{su13} (note 
that SAAO - Landolt in their figure). There is no significant 
difference in $V-I$ up to 3.5, while redder stars from 
the CTIO data are slightly bluer than those of from HSB12, which 
is very similar to the trend shown in \citet{be95}.

The dashed lines in the figures trace the mean differences 
between the two sets of data in $V$ and $I$ bands. After 
correcting for these mean differences, the $V$ and $I$ magnitudes 
of the CTIO data were transformed to the SAAO system. HSB12 
did not obtain the $R-I$ colors of stars, and therefore we 
could not directly check the systematic difference between 
the SAAO and Landolt systems. However, it is known that there 
is no significant difference in $R-I$ between the two standard 
systems. $R-I$ colors transformed to the Landolt system were 
used in this study.

For bright stars ($V\leq 13.5$ mag), 
the two sets of data were combined into a single photometric 
catalogue by computing the weighted-mean magnitudes and colors, 
where the inverse of squared error was used as the weight value. 
For faint stars ($V > 13.5$ mag), we only used the CTIO data 
($V$, $V-I$, and $R-I$) because the spatial resolution of 
the new data is better than that of HSB12. All sources from HSB12 
without any counterparts in the CTIO data were visually checked 
in both images from SSO and CTIO, of which 38 sources were rejected 
because they are very faint ($V>18$ mag).

\begin{figure}[ht]
\centering
\includegraphics[width=0.8\columnwidth]{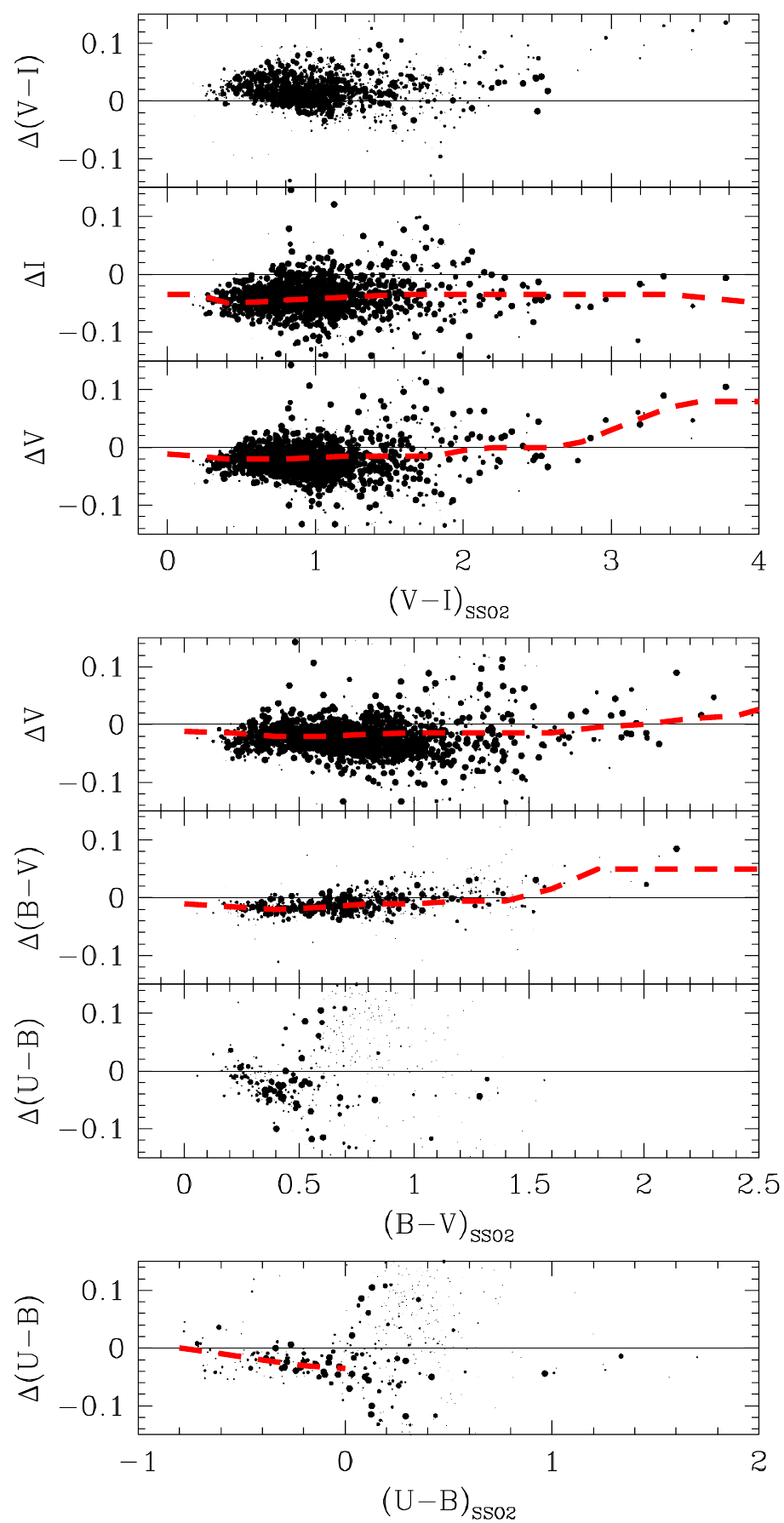}
\caption{Comparison of photometry -- SSO2 minus HSB12.
The symbols are the same as in Figure~\ref{comp_4m_sso}.}
\label{comp_sso_Tr 15}
\end{figure}

We cross-matched the SSO2 data (the Landolt system) with 
the combined CTIO-HSB12 data (the SAAO system). From a visual inspection 
of stars in SSO2 without any counterpart in the combined data, 
we rejected 12 spurious sources from SSO2. Figure~\ref{comp_sso_Tr 15} 
shows the systematic difference for stars with $V\leq$ 17.0 mag. 
Overall, the systematic difference between two data sets is very 
similar to the comparison in \citet{su13}. The difference in $V$ 
is very similar to the pattern of the systematic difference between the 
CTIO and HSB12 data, while the difference in $I$ shows only zero-point 
difference without any color dependency, and the combination 
of both differences results in the systematic difference in $V-I$. 
We transformed the SSO2 data ($V$, $I$, and $B-V$) to the SAAO 
system by correcting the mean differences (dashed lines in Figure~\ref{comp_sso_Tr 15}). 

In $U-B$, there is a non-linear difference between the data 
in HSB12 and SSO2. Such a non-linear 
pattern comes from the $U-B$ colors of the Landolt standard 
stars for A-type stars because the transmission curve of 
his $U$ filter is severely affected by the Balmer discontinuity 
\citep{su00}. For this reason, the $U-B$ colors of A-type 
stars transformed to the Landolt's system (SSO2) could not 
properly be transformed to the $U-B$ colors of the SAAO system. 
In addition, faint stars with $U-B \geq -0.1$ show a large scatter 
due to the large photometric errors in $U$ band. Hence, we did 
not use stars with $U-B$ colors redder than $-0.1$ in transformation 
to the SAAO system (see the dashed line in the bottom panel 
of Figure~\ref{comp_sso_Tr 15}.) We combined the SSO2 and 
CTIO-HSB12 data into a master photometric catalogue 
(Table~\ref{table_data}) in the same manner as above.

\begin{figure}
\centering
\includegraphics[width=0.8\columnwidth]{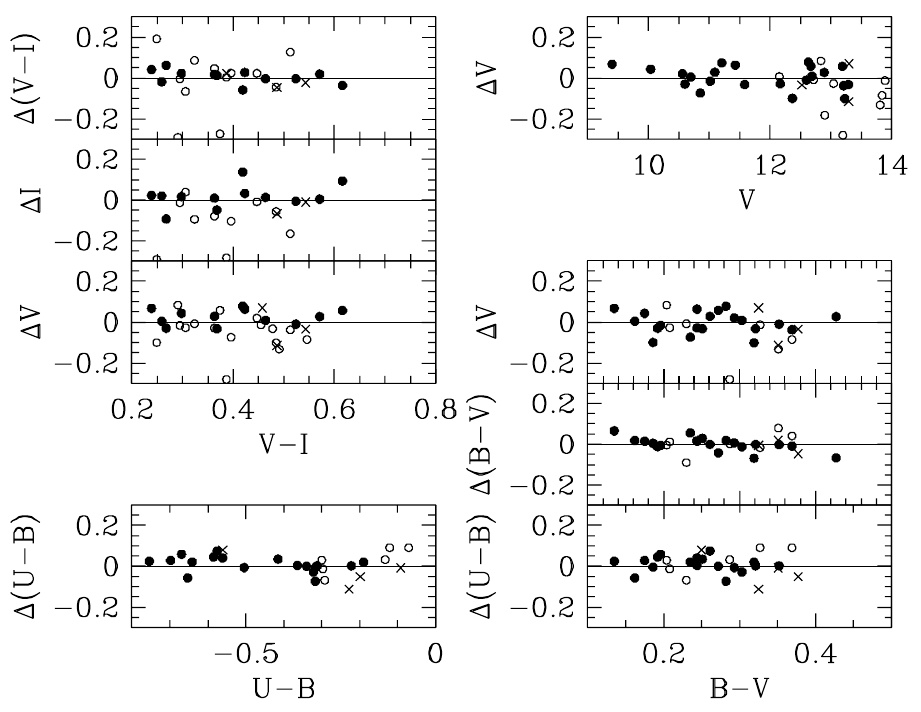}
\caption{
Comparison with \citet{fe80} of photometry for the stars in Tr 15 .
Crosses represent variables in \citet{fe80} or photometric doubles in our data.
Open circles are stars with a single measurement in \citet{fe80} or $\epsilon>0.05$ mag in our data.
Filled circles represent the stars used in the calculation of mean photometric differences (see the text).}
\label{comp_f80}
\end{figure}

The comparison of our data with the photoelectric 
photometry of \citet{fe80} for Tr 15 shows a good consistency (Figure~\ref{comp_f80}) [$\Delta V=0.003\pm 0.054$ (22 stars); $\Delta I= 0.022\pm0.060$ (13 stars); $\Delta(U-B)=0.010\pm0.037$ (18 stars); $\Delta(B-V)=0.000\pm0.034$ (22 stars); $\Delta(V-I)=0.003\pm0.034$ 
(13 stars)]. Variables from table 1 of \citet{fe80} were excluded in comparison. Also, we did not use several photometric 
double stars that were spatially not resolved in photometry. In some 
cases, these photometric double stars can be resolved by PSF 
photometry, while it is impossible to resolve such stars using 
a simple aperture photometry conducted with photoelectic observations. 
The photometric data of these photometric double stars are not 
suitable for comparison. The comparisons of photometry with other 
previous studies were summarized in HSB12. 

\begin{landscape}
\begin{table*}[ht]
\caption{The Counterparts of Optical Sources in the Previous Surveys}
 \label{table_cid}
\makebox[0.8\linewidth][c]{
  \begin{tabular}{ccccccccc}
\hline
\hline
 ID   & 2MASS$^{1}$     &Vela-Carina$^{2}$& CCCP$^{3}$       & VISTA$^{4}$      &HD$^{5}$   &CPD$^{6}$  &ALS$^{7}$& Gaia EDR3$^{8}$   \\
\hline
 24977&                 &G287.4437-00.8251&                  &104315.51-594707.2&           &           &         &5350354530047696128\\
 38714&10434743-5931264 &G287.3798-00.5629&                  &104347.43-593126.5&           &           &         &5350408680997497600\\
 44874&                 &                 &                  &104400.80-591710.5&           &           &         &5350360577348417664\\
 49408&                 &                 &                  &104410.84-593950.6&           &           &         &                   \\
 61317&10443591-5923356 &G287.4091-00.3992&104435.91-592335.7&                  &           &           &         &                   \\
 62149&10443766-5923073 &G287.4087-00.3905&104437.69-592307.2&104437.67-592307.2&           &           &         &5350395486857958912\\
 62676&10443864-5930076 &G287.4652-00.4928&                  &104438.66-593007.8&           &           &         &5350357308862763520\\ 
 70049&10445473-5944032 &G287.6036-00.6822&                  &104454.73-594403.2&           &           &         &5350358000381591936\\
 71927&                 &G287.6523-00.7563&                  &                  &           &CPD-59 2617&         &                   \\
 85757&                 &                 &                  &104528.59-592646.4&           &           &         &                   \\
 91223&10453965-5940079 &G287.6567-00.5804&                  &                  &           &           &         &5350335048073756544\\ 
127443&10465166-5926443 &G287.6883-00.3122&                  &104651.64-592644.3&           &           &         &5350344733197719424\\
\hline
 \end{tabular}
}
\begin{tabular}{l}
This table is available in its entirety in the online journal.
Only a portion is shown here for guidance regarding its form and content. \\
$^{1}$ The $JHK_S$ NIR photometric data \citep{sk06}. \\
$^{2}$ The {\it Spitzer} MIR photometric data \citep{ma07}. \\
$^{3}$ The CCCP X-ray point source catalog \citep{br11b}. \\
The HWAK-I identification for the $JHK_S$ magnitudes \citep{pr11} are also available in this catalog. \\
$^{4}$ The VISTA $JHK_S$ NIR photometric data \citep{pr14}. \\
$^{5}$ The Henry Draper catalog and Henry Draper Extension catalog number. \\
$^{6}$ The Cape Photographic Durchmusterung catalog number. \\
$^{7}$ The Alma Luminous Star catalog number \citep{als}.\\
$^{8}$ The Gaia EDR3 \citep{edr3}.\\
 \end{tabular}
\end{table*}
\end{landscape}

\subsection{Completeness Test\label{completeness}}
The completeness of photometry is, in general, limited 
by the bright nebulae, high-stellar density, and very 
bright stars. In particular, the Carina Nebula is one of 
the brightest nebulae in the sky and is visible to the naked 
eyes. The brightness of the central part (around Tr 14 and 
Tr 16) is mostly $\sim10$\% and partly over $20$\% of the 
saturation level in the long exposed $V$ images (in the 
extreme case it reaches up to $\sim50$\% in the long 
exposed $R$ images near Tr 14). The nebula around Cr 232 is also bright, 
about 75\% level of the nebulocity in the region around Tr 14 and Tr 16.
Bright stars in Tr 14, Tr 15, and Tr 16 
also have considerable influence on the completeness of the 
photometry because some faint stars 
can be located in their intense, extended wings.

We performed completeness test using artificial stars 
generated to be distributed randomly in magnitude but 
at known positions. Artificial images were generated on 
short- and long-exposed $V$, $R$, and $I$ images after 
subtracting all known stars. A total of 9,248 
artificial stars were added onto 1024$\times$2048 pixel$^2$ 
area in 15-pixel intervals in both CCD X and Y coordinates. 
Then we performed the same reduction procedures as described 
in Section~\ref{ctio4m} including visual inspections, and 
then transformed to the standard system.

\begin{figure*}
\includegraphics[width=0.69\columnwidth]{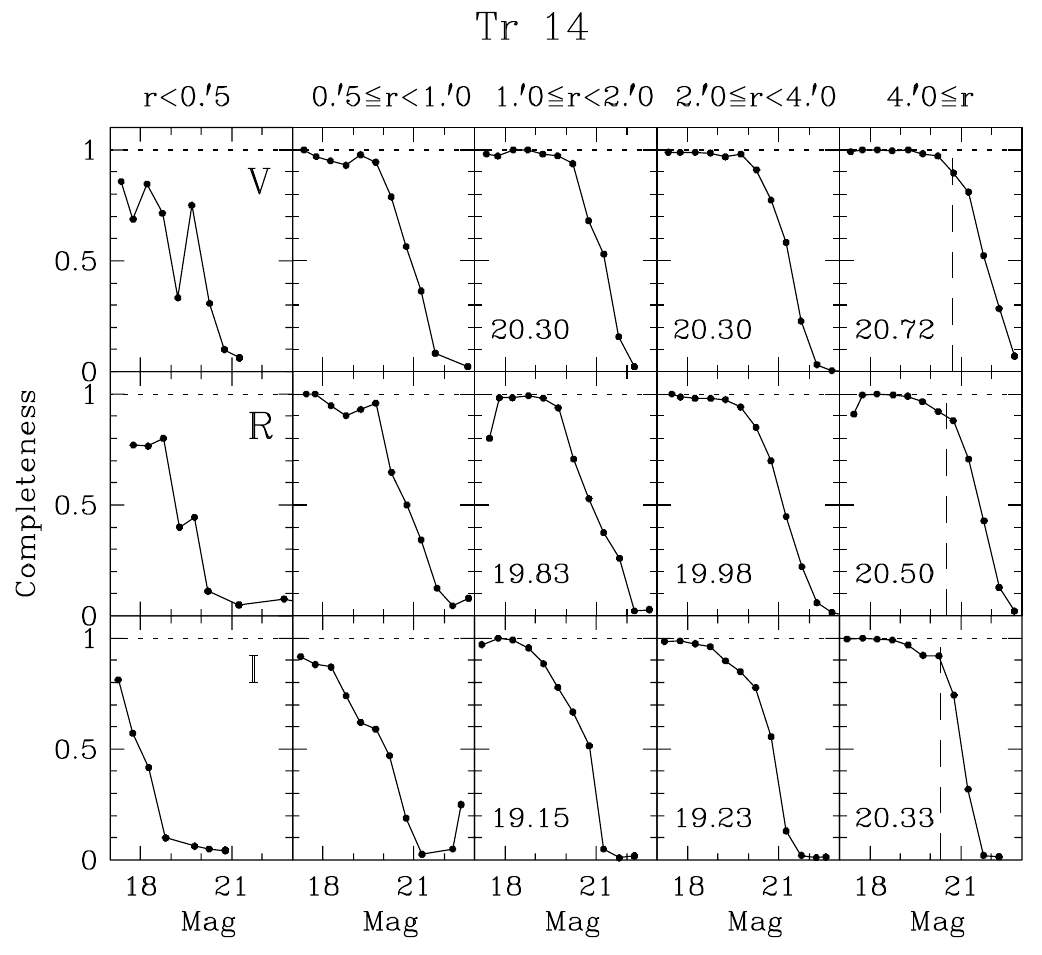}
\includegraphics[width=0.69\columnwidth]{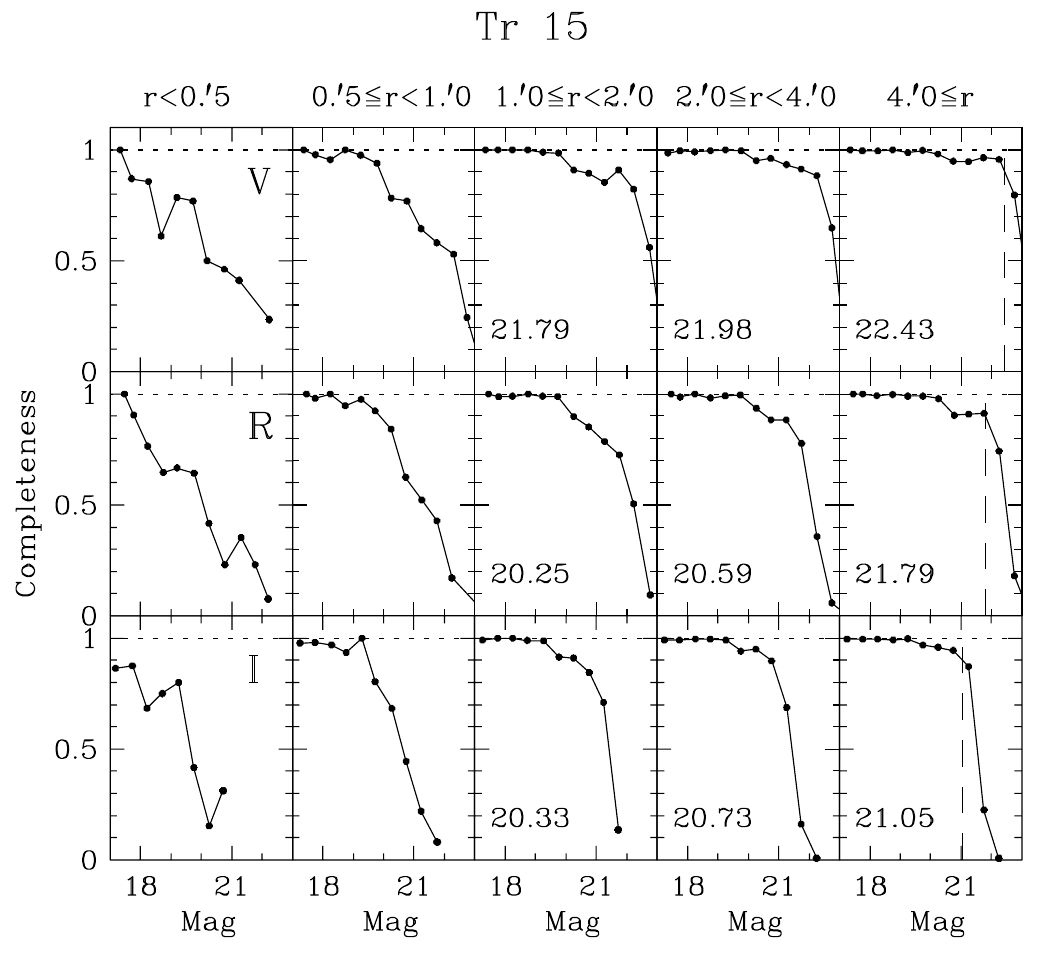}
\includegraphics[width=0.69\columnwidth]{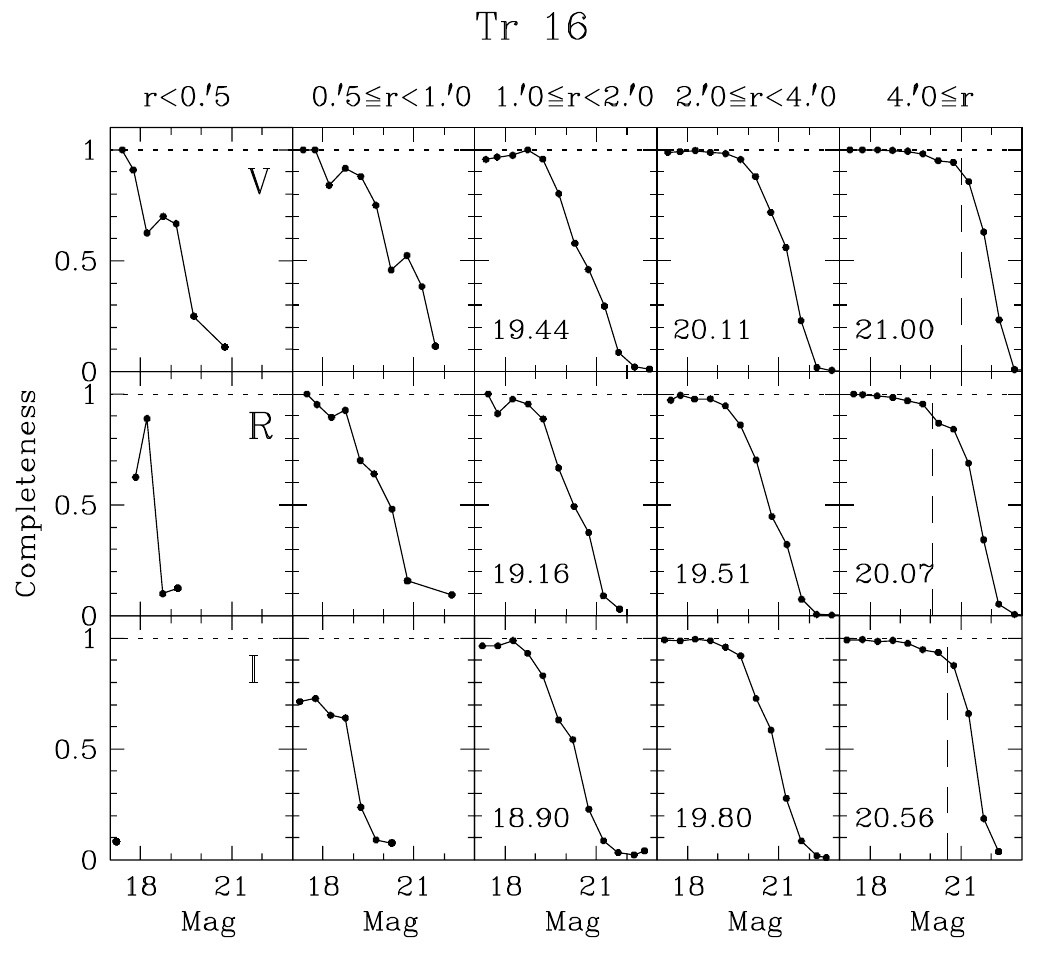}
\caption{
The completeness of our photometry at different radial distances from the brightest stars of each cluster.
The numbers in the boxes are the magnitudes of the 90\% completeness limit (the vertical dashed line).}
\label{completeness_fig}
\end{figure*}

Figure~\ref{completeness_fig} displays 
the completeness levels of our photometry with respect to the 
projected distances from the brightest stars in the clusters 
(HD 93129 for Tr 14, HD 93249 for Tr 15, and $\eta$ Carinae for Tr 16). 
The completeness of stars in the innermost region is 
highly affected by the extended wings 
of the saturated stars. The completeness is progressively 
improved in the outer rings ($r\geq 1^{\prime}$ ). The 90\% 
completeness limits in $V$ at $r\geq 4.0^{\prime}$ are found at $20.7$, 
$22.5$, and $21.0$ mag for Tr 14, Tr 15, and Tr 16, respectively. 
Tr 15 region shows a better completeness level since the
nebulosity around this cluster is weaker. The effect of the nebulosity on the 
completeness is particularly severe in $V$ and $R$, while $I$ 
images are less affected by the nebulosity. The $I$ band 
photometry is thus more useful for the identification 
of faint sources. The completeness limit in the outer regions 
of Tr 14 and Tr 16 showing the high nebulocity is similar in $V$ and $I$ although the $V-I$ color of stars near the completeness 
limit is $V-I\gtrsim1.5$, while the limiting magnitude in $V$ 
is more than one magnitude fainter than that in $I$ in the 
outer region of Tr 15. The limiting magnitudes in the Cr 232 region are slightly fainter than those in the outer regions of Tr 14 and 16, given the different brightness of surrounding nebulae.

\section{Cross-Identification of Stars in the Previous Surveys}
Multi-wavelength data are very useful for member 
selection and studying the properties of star-forming regions. There 
are several extensive surveys to study young stars in the Carina Nebula. 
In this section, we describe the cross-identification of our optical data with 
the previous large surveys, e.g., the CCCP in X-ray \citep{to11}, the 2MASS 
and VISTA in NIR, the Vela-Carina survey in MIR wavelengths, and the Gaia 
early data release 3 (EDR3; \citealt{edr3}).
The results from cross-matches 
between the optical data and the previous surveys are shown in Table~\ref{table_cid}.    
We also identified the bright stars in the HD, CPD, and ALS catalogs, and 
listed them in the table.

\subsection{The Chandra Carina Complex Project (CCCP)}
X-ray emission is of typical properties of PMS stars and gives 
the most reliable selection criterion for PMS members in young open 
clusters \citep{su04,wo05}. Many O-type stars in the Carina Nebula 
are also strong X-ray emission stars \citep{ga11,po11b}. We searched 
for the optical counterparts of the CCCP sources \citep{to11,br11b} 
within two different matching radii of $0.8^{\prime\prime}$ -- 
$2.0^{\prime\prime}$ given the position error of the CCCP survey. 
Within the matching radii, the closest optical 
source was assigned as the X-ray emission star. A total of 4,229 optical 
sources were identified as X-ray emission stars.

\citet{br11a} divided the X-ray sources into four classes, e.g.,  ``H1'' 
(foreground stars), ``H2'' (young stars in the Carina Nebula), ``H3'' 
(background stars), ``H4'' (extragalactic sources), and unclassified 
stars. Within the FOV of our optical photometry, the detection rates 
of the CCCP sources in optical passbands were 49.3\% (73/148), 73.0\% 
(4,046/5,546), 100\% (1/1), 8.3\% (11/132), and 31.8\% (98/308) 
for ``H1'', ``H2'', ``H3'', ``H4'', and unclassified sources, 
respectively. The fraction of optically detected 
extragalactic sources (H4) is very small. These extragalactic 
sources are expected to be highly obscured, especially in the 
optical passbands, by the giant molecular clouds behind the 
Carina Nebula. In addition, their surface brightness is too 
low to be detected in bright emission nebulae.

It is noticeable that the optical detection rate of ``H1'' 
sources (49.3\%) is lower than that of ``H2'' sources (73.0\%). 
The simulation for the source contamination in the CCCP 
observation \citep{ge11} suggests that the observed median 
X-ray energy of many ``H1'' sources is slightly lower than 
1 keV. The model distribution for $J$-band flux of the 
simulated ``H1'' objects is similar to or fainter than that 
of ``H2'' objects and is bimodal in shape. These facts imply 
that the optically undetected ``H1'' sources are mostly 
foreground M-type dwarfs. 

In order to understand the lower optical detection rate 
of ``H1'' sources, we simply simulated the detection rate 
of foreground M-type dwarfs in both the optical $I$ band 
and X-ray. For the foreground stellar distribution, we 
adopted the synthetic stellar population for the Galactic 
disk model \citep{ro03,ro04}. The distribution of M-dwarf 
stars in the foreground was generated for $d=0$ -- $2.4$ kpc 
and $E(B-V)=0.15$/kpc (see Sections~\ref{rv_var} and ~\ref{dist} 
for the reddening and distance). For the 90\% completeness 
limit of $I\leq 21.1$ mag around Tr 15, 59.9\% of the M-dwarfs 
are expected to be detected in $I$. If we assume $\log (L_X/L_{Bol}) = -3$ \citep{vi87,wr11}, 0.37\% and 0.26\% of M-dwarfs are expected 
to be detected in X-ray and in both $I$ and X-ray, respectively. 
The simulated optical detection rate of the stars detected 
in X-ray is 68.9\% for foreground M-dwarfs. For the Tr 14 and 
Tr 16 fields, the detection rate decreases down to 55.7\%, which is close 
to the optical detection rate of ``H1'' sources (49.3\%). Since 
we adopted the saturation level of $\log (L_X/L_{Bol})$ for 
the simulation, these detection limits are upper limits. We 
conclude that the major reason of the lower optical detection 
rate of ``H1'' sources in the CCCP catalog is their intrinsic 
faintness in the optical bands.

\subsection{$JHK_S$ Near-Infrared Data\label{nir}}
NIR photometric data are very useful 
to detect not only cool stars and active YSOs with 
circumstellar disks because these stars emit most energies 
through their photospheres in NIR passbands. The effects 
of interstellar extinction in infrared are much smaller 
than those in optical passbands. In addition, the various color excess 
ratios derived from optical-NIR colors such as $V-J$, $V-H$, 
and $V-K_S$, allow us to examine the reddening law toward 
star-forming regions. Such optical-NIR colors have wide 
baselines, and therefore it is also possible to discriminate 
the bona-fide members of a star-forming region from foreground 
or background field stars (see the CMDs in \citealt{hur15}). 

In order to take these advantages, 
we identified the optical counterparts of $JHK_S$ NIR sources 
from the VLT HWAK-I data \citep{pr11} and the VISTA survey 
data \citep{pr14}. The HWAK-I data provides $JHK_S$ magnitudes 
for the CCCP sources. Their limiting 
magnitudes are slightly fainter than those of the VISTA data. 
The NIR data of the optical sources with the counterparts CCCP 
survey were obtained from the HWAK-I $JHK_S$ data, while those of 
the other sources were taken from the VISTA data. We took the 
2MASS data \citep{sk06} for some bright sources without the 
counterparts of the VISTA data and HWAK-I data. A total of 
114,458 (84.7\%) optical sources have NIR counterparts (109,102 
from VISTA, 3,585 from HWAK-I, and 1,771 from 2MASS). If we exclude 
the optical sources detected only in $I$, 85,450 out of 95,292 
stars (89.7\%) with $V-I$ or $R-I$ color have NIR 
data. 

\subsection{The {\it Spitzer} Vela-Carina Survey\label{mir}}
MIR data are very useful in identifying PMS stars with accretion 
disks \citep{L87}. The Vela-Carina survey \citep{ma07} published 
the MIR point source catalog for four {\it Spitzer}/IRAC bands 
(3.6, 4.5, 5.8, and 8.0 $\mu$m) over {\it l}$=255^{\circ}$ 
-- $295^{\circ}$ along the Galactic plane. We identified the 
optical counterpart of the MIR sources in the Vela-Carina survey 
catalog (more reliable list with reliability $> 99.5$\%)\footnote{\: http://irsa.ipac.caltech.edu/data/SPITZER/GLIMPSE/doc/velacar\_dataprod\_v1.0.pdf} within a matching radius of $1.0^{\prime\prime}$. We found a total of 15,748 
stars in common between the optical and MIR data. 

\begin{figure}
\includegraphics[width=\columnwidth]{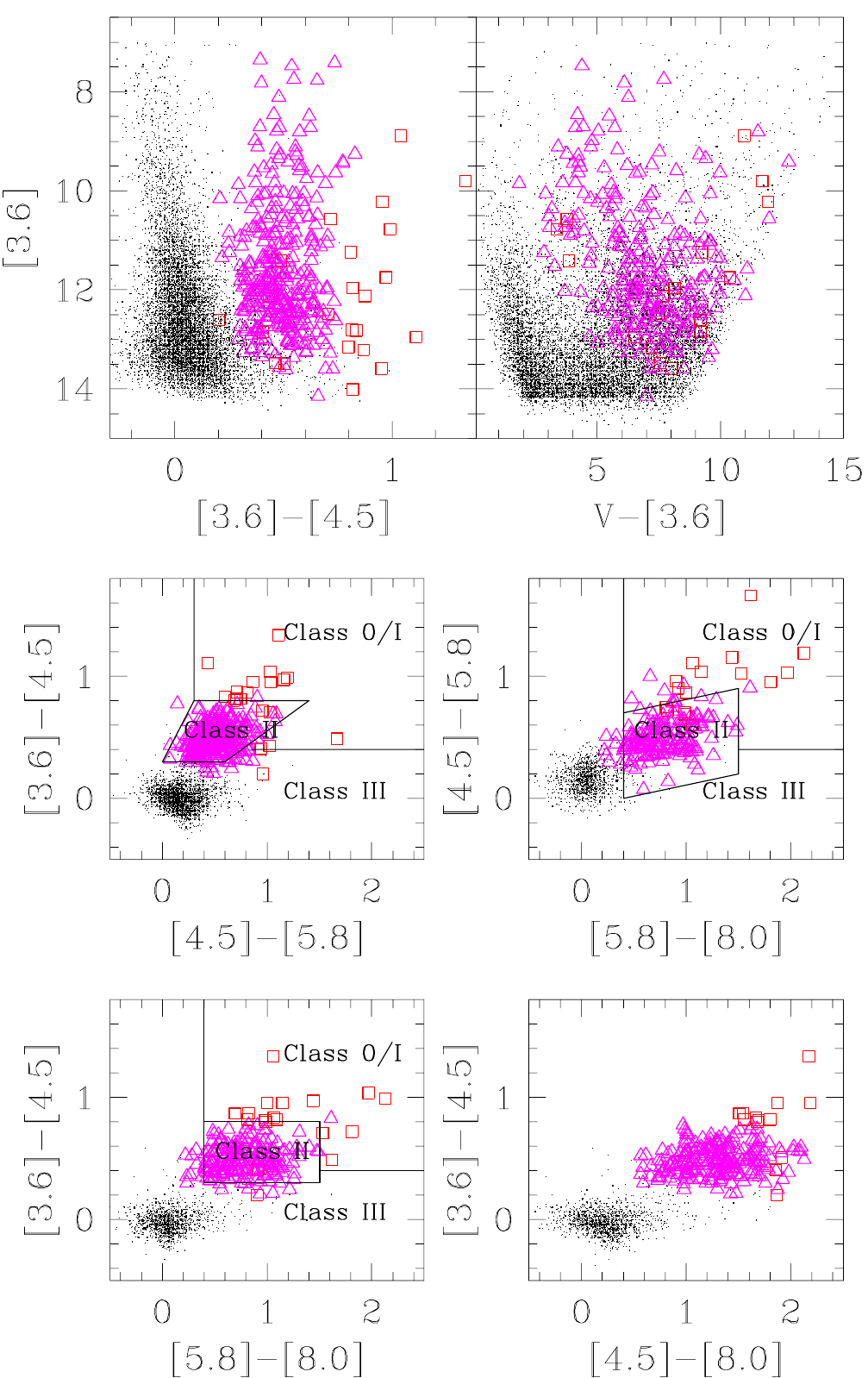}
\caption{
YSO classification of the optically detected MIR sources in the Vela-Carina catalog.
Squares (red), triangles (magenta), and dots (black) represent class 0/I, class II, and class III stars, respectively.}
\label{class_mir}
\end{figure}

We used the MIR data to classify young stellar objects (YSOs) 
into three evolutionary stages, e.g., Class 
0/I with dusty envelopes and 
disks, Class II with disks, and Class III with photospheric colors. The 
classification scheme of \citet{hur15} was used (see also \citealt{su09}).
The CMD and TCDs of the MIR sources with optical counterparts are shown 
in Figure~\ref{class_mir}. A total of 26 class 0/I, 341 class II, and 
2,514 class III stars were identified.

\subsection{Gaia Early Data Release 3}
The Gaia satellite was launched in 2013 to measure the parallaxes and 
proper motions of more than a billion stars in the Galaxy \citep{gaia}.
We used Early Data Release 3 (EDR; \citealt{edr3}) in this study. There are 
a total of 91,168 Gaia EDR3 sources in the FOV of our data, and 59,460 
out of the sources were matched with the stars in our optical photometric 
data using a matching radius of $0.5^{\prime\prime}$.

It was reported that there are systematic zero-point offsets in the parallaxes 
of EDR3 \citep{li21}.  Such systematic offsets depend on magnitude, color, 
and ecliptic latitude. We corrected parallaxes for the zero-point 
offsets according to the recipe of \citet[\url{https://gitlab.com/icc-ub/public/gaiadr3_zeropoint}]{li21}. 
Stars with renormalised unit weight error (RUWE)$\leq 1.4$ 
and parallax-to-error ratio $> 5$ were used in analysis.

\subsection{The Gaia-ESO Spectroscopic Survey}
\citet{da17} presented the spectroscopic survey data performed 
for the identification of young stars in the Carina Nebula. They 
obtained the spectra of 1,085 stars with $V=$ 12--19 mag using VLT-FLAMES/Giraffe, and selected 298 young stars based mainly 
on the presence of lithium absorption, radial velocity 
(RV), H$\alpha$ emission, and the X-ray emission from the CCCP data.

As lithium absorption and H$\alpha$ emission are reliable criteria 
for selection of PMS stars in young open clusters, a total 
of 285 stars that have equivalent width (EW) measurement of lithium 
absorption or detection of H$\alpha$ emission (hereafter LH stars) 
were selected. We did not use RV data in our membership 
selection because there is a small separation between the RV 
distributions of members and foreground stars (see figure 3 in \citealt{da17}). 

\section{Reddening}
At optical wavelengths, the relation between total extinction 
in $V$ ($A_V$) and reddening in $B-V$ [$E(B-V)$] is expressed 
as $R_V$, the total-to-selective extinction ratio, i.e. 
$A_V=R_V\times E(B-V)$. $R_V=3.1$ is, in general, adopted as 
the normal reddening law \citep{gu89,dr03,su14}, while some 
young open clusters or star-forming regions show a higher 
$R_V$. It was also reported that the Carina Nebula shows 
abnormal reddening law expressed by two components, the foreground reddening and the intracluster reddening, (\citealt{he76,fo78,sm87,va96}; 
HSB12), i.e., 
\begin{equation}
A_V= 3.1\times E(B-V)_{fg} + R_{V,cl}\times E(B-V)_{cl}
\end{equation}
\noindent where $R_{V,fg}$, $E(B-V)_{fg}$, $R_{V,cl}$, and 
$E(B-V)_{cl}$ represent the foreground $R_V$, foreground 
reddening in $B-V$, and $R_V$ and $E(B-V)$ of the intracluster 
medium, respectively. Several previous studies determined 
the values of $E(B-V)_{fg}$ and $R_{V,cl}$ to be 0.2 to 0.3 
and 4.3 to 4.7, respectively \citep{he76,fo78,va96}. HSB12 
inferred the reddening law toward the Carina Nebula to be 
$R_{V,cl}=4.4\pm0.2$ with $E(B-V)_{fg}=0.36\pm0.04$.

HSB12 adopted the reddening vector $E(U-B)/E(B-V)=0.72$ to 
determine the reddening values of individual early-type stars. 
However, an additional term associated with 
reddening values, in fact, has additional influence on the 
reddening vector of highly reddened OB stars ($E(U-B)/E(B-V)=0.72+0.025E(B-V)$ --  \citealt{sos0,hur15}. 
In this study, we revisit the properties of reddening over 
the Carina Nebula with the modified reddening vector.

\begin{figure}
\includegraphics[width=\columnwidth]{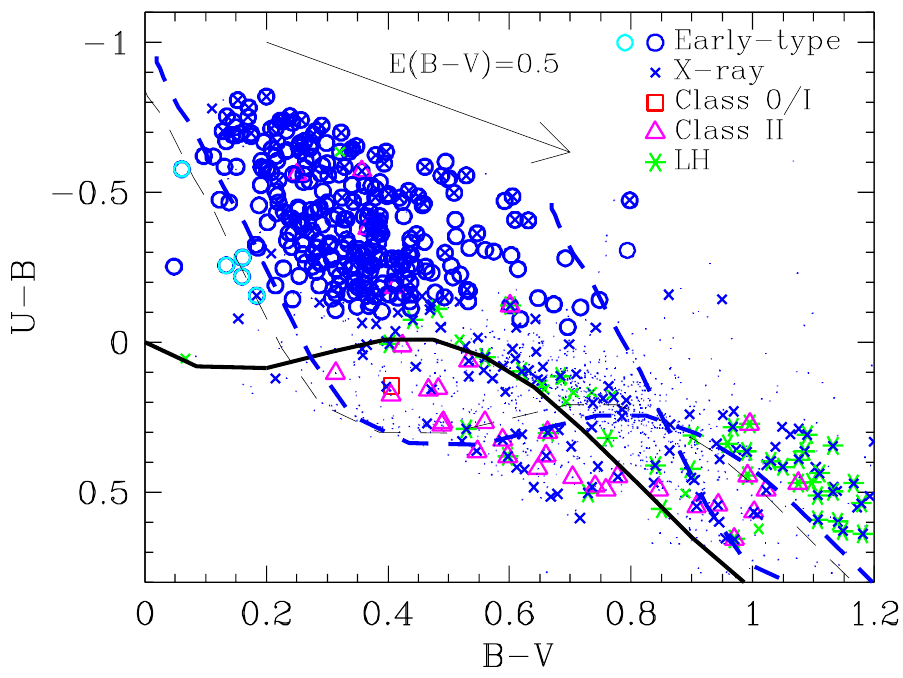}
\caption{
($U-B, B-V$) diagram for $V\leq17.0$ mag.
The solid (thick black) line is the unreddened ZAMS relation while dashed lines represent
the reddened ZAMS relations for $E(B-V)=0.30$ (thin black), $0.35$ (thick blue) and $1.0$ (thick blue), respectively.
The selected early-type stars are shown as blue circles, while cyan circles are the foreground early-type stars at $E(B-V)\sim0.3$
(see Section~\ref{dist}).  
X-ray emission stars, class 0/I, class II, and others, respectively.
The asterisks (green) represent the LH stars(lithium absorption or H$\alpha$ emission from \citealt{da17}).}
\label{ubbv}
\end{figure}

\subsection{Early-Type Stars and Their Reddening\label{sel_er}}
The reddening across star-forming regions can be traced by 
using early-type stars because their reddening values 
can be determined in ($U-B,B-V$) TCD along the reddening vector. 
Stars with a spectral type of B3 or earlier are considered as 
early-type stars. For O-type stars, we adopted spectral 
classifications from \citet{so14}. For some O-type stars 
not listed in \citet{so14} and B-type stars, we adopted the 
spectral classifications from other previous studies  \citep{fe80,le82,le91a,le91b,mo88,mj93,ga11,sk14,ka15,da17}. 
A list of 116 early-type stars was obtained from the 
data bases of spectral classification.

Figure 5 in HSB12 shows that there are many probable 
early-type stars without a spectral type. Early-type stars are bright and have blue colors, and therefore 
they can be identified by using their photometric properties. 
The reddening-free parameter Johnson's Q parameter 
[$\equiv (U-B)-0.72(B-V)$ -- \citealt{JM53}] is of useful 
tool to discriminate early-type stars. Smaller Q values 
indicate earlier types stars. $Q \sim -0.5$, in general, 
corresponds to the spectral type of mid-B \citep{JM53}.
So, stars with $V< 15.0$ mag and Johnson's $Q$ [$=(U-B)-0.72(B-V)$] $\leq -0.5$, 
or with $V<15.0$ mag, $U-B<-0.1$, and $B-V<0.45$ were selected as early-type 
stars. We did not use fainter stars ($U>17.0$ or $B>17.0$ mag) 
or stars with large photometric 
error ($\epsilon_Q = \sqrt{\epsilon_{(U-B)}^2+\epsilon_{(B-V)}^2} > 0.1$). 

These criteria are very similar to those of HSB12. Several 
highly reddened ($E(B-V)>0.7$) stars clearly 
showed $R_V$ close to the normal value of 
$R_V=3.1$ (see figure 5 of HSB12). If these stars were real 
early-type stars toward the Carina Nebula, they would be 
reddened by the same interstellar media along the line of 
sight as the other early-type stars, and should show 
the same reddening law. Therefore, these stars were suspected 
to be foreground FG-type stars, and we excluded them from the 
early-type star list. A total of 186 additional early-type 
stars were selected from our photometric data.

We confirmed, as already noted in HSB12, that there are a 
few highly reddened [$E(B-V)>1.0$] early-type candidates 
selected from our criteria Figure~\ref{ubbv}. One of them, 
ID 50468 (ALS 15210), is an O3.5 If* Nwk star classified 
by \citet{so14} and shows strong X-ray emission [Net counts 
(0.5--8 keV) = 359.2, \citealt{br11b}], but the others are 
quiescent in the X-ray. ID 62439 shows weak X-ray emission and 
was classified as a young star (``H2'') in the Carina Nebula 
by \citet{br11a}. Active PMS stars have 
blue colors particularly in $U-B$ because the energy 
generated by mass accretion emits in ultraviolet 
wavelength \citep{RHS00,LSK14,li14}. Therefore, in the 
($U-B$, $B-V$) diagram, such PMS stars overlap with highly 
reddened early-type stars. Spectroscopic observations are 
required to identify these stars. In this study, 
we cautiously excluded these stars from our early-type star list.
ID 11820 is fainter but bluer ($V=14.680$, $U-B=-0.995$, $B-V=-0.205$) than 
other early-type stars. This star could be a foreground sdB star.
The selected early-type stars are shown in Figure~\ref{ubbv}.
\begin{figure}
\includegraphics[width=\columnwidth]{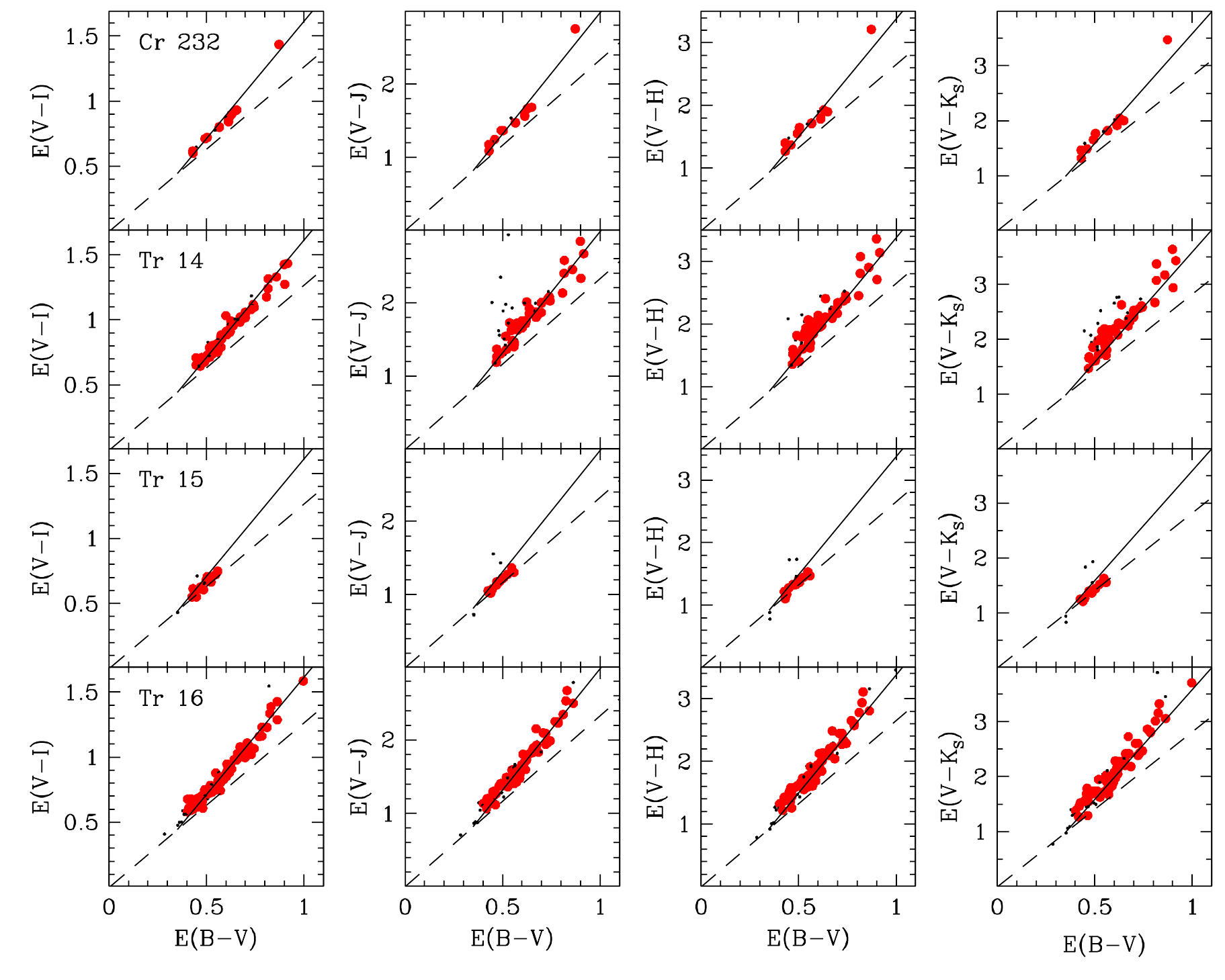}
\caption{
[$E(V-\lambda)$, $E(B-V)$] diagrams for $V-I$, $V-J$, $V-H$, and $V-K_S$.
The solid and dashed lines correspond to $R_{V,cl}=4.4$ and $R_V=3.1$, respectively. 
The red dots represent the stars used to plot the $R_{V,cl}$ variation map,
while the small dots are the stars excluded in analysis.}\label{rv_cl}
\end{figure}

\begin{figure}[ht]
\includegraphics[width=\columnwidth]{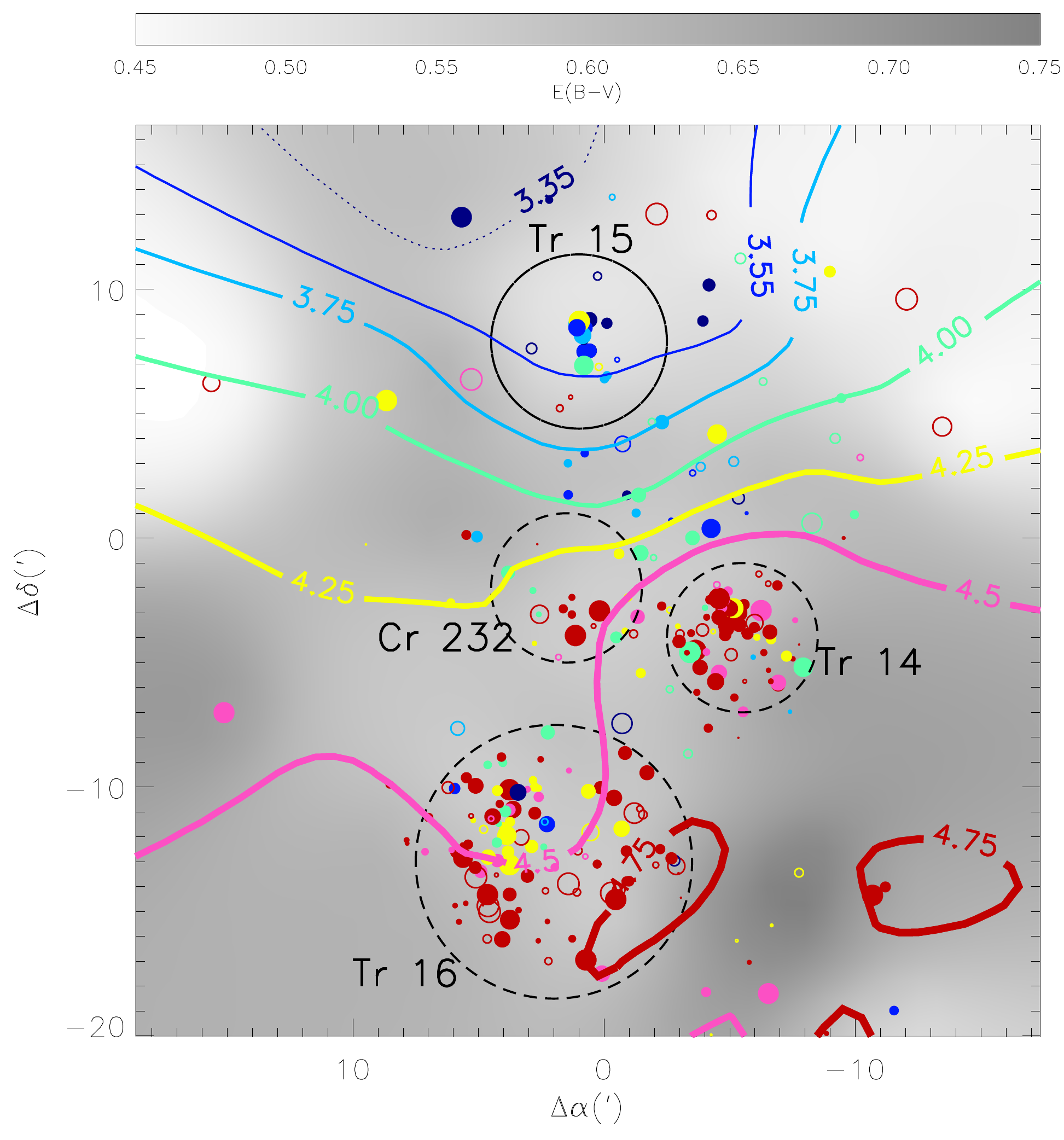}
\caption{
Spatial variation of $R_{V,cl}$ and $E(B-V)$.
The contours represent $R_{V,cl}=3.35$ (black dashed), $3.55$ (blue), $3.75$ (cyan), $4.0$ (green), $4.25$ (yellow), $4.5$ (magenta) , and $4.75$ (red).
The color of stars indicates their derived $R_{V,cl}$ values.
Filled circles are the stars used in the calculation of the contours, while open circles are the stars excluded due to their probable NIR excess,
small $E(B-V)$ ($<0.4$ mag), photometric duplicity in the optical data, or variability reported by previous studies.
The size of dots is proportional to the brightness of the stars.
The grey density map shows the spatial variation of $E(B-V)$ from the early-type stars without the known variability and the photometric duplicity.}
\label{rv_spt}
\end{figure}

We determined the reddening values of individual stars along the 
modified reddening vector \citep{sos0} in Figure~\ref{ubbv}. The 
mean reddening is about 0.58 mag, which is close to that obtained 
by HSB12(0.61 mag). It was also confirmed that there is a 
high-level of differential reddening up to 1.28 mag. We obtained 
the intrinsic colors, such as $(V-I)_0$, $(V-J)_0$, $(V-H)_0$, 
and $(V-K_S)_0$, interpolating the reddening-corrected $B-V$ 
colors to their intrinsic color relations with respect to $B-V$ 
\citep{sos0}, and then their color excesses were 
computed by comparing the observed colors with the intrinsic 
colors.

As noted in HSB12, there are several foreground early-type stars 
($E(B-V)\sim 0.3$) toward the Carina Nebula. Indeed, five of 
them are $\sim 2.0$ kpc away from the Sun, according to the 
Gaia parallaxes \citep{edr3}. The distances of these stars 
are smaller than the known distance to the Carina Nebula. 
These stars were used to constrain the 
foreground reddening and examine the foreground reddening law. 

\subsection{Spatial Variation of $R_{V,cl}$ \label{rv_var}}
$R_V$ was derived using various color excess ratios by means of the 
Equations 5 -- 8 in \citet{sos0}. These color combination 
allows us to exam the reddening law in a consistent way over a wide 
wavelength range.

\begin{table*}
\centering
\caption{
The Averaged $UBVI$ Data from the Previous Photometric Studies for the Bright Saturated Stars } 
 \label{star_sat}
\begin{tabular}{lcccc|lcccc}
\hline
\hline
Star & $V$ & $U-B$ & $B-V$ & $V-I$ & Star & $V$ & $U-B$ & $B-V$ & $V-I$ \\
\hline
HD 93129A  &  7.16  & -0.80  & 0.21  & 0.44 & HD 93190   &  8.58  & -0.82  & 0.33  & 0.63 \\
HD 93403   &  7.30  & -0.80  & 0.22  & 0.39 & Tr 16 - 100&  8.60  & -0.78  & 0.22  & 0.52 \\
HD 93250   &  7.37  & -0.85  & 0.17  & 0.37 & HD 93161B  &  8.60  & -0.77  & 0.23  & 0.42 \\
HD 93205$^{a}$   &  7.75  & -0.91  & 0.05  & 0.23 & V 572 Car$^{a}$  &  8.78  & -0.80  & 0.14  & 0.31 \\
HD 93160   &  7.82  & -0.78  & 0.17  & 0.36 & HD 93128   &  8.79  & -0.77  & 0.25  & 0.43 \\
HD 93130$^{a}$   &  8.06  & -0.77  & 0.22  &        & HD 93129B  &  8.84  & -0.79  & 0.23  & 0.46 \\
HD 93162$^{a}$    &  8.10  & -0.69  & 0.41  & 0.96 & HDE 303311  &  9.05  & -0.86  & 0.13  &        \\
HDE 303308  &  8.17  & -0.87  & 0.13  & 0.30 &HDE 303304  &  9.67  & -0.69  & 0.40  & 0.74 \\
HD 93161A  &  8.31  & -0.73  & 0.21  & 0.38 & HDE 303312  &  9.97  & -0.66  & 0.32  &        \\
HD 93249   &  8.36  & -0.75  & 0.14  & 0.30 & HDE 303302$^{a}$  & 10.79  & -0.51  & 0.20  & 0.35 \\
HD 93204   &  8.43  & -0.89  & 0.09  & 0.32 \\ 
\hline
\end{tabular}
\begin{tabular}{p{5.1in}}
The magnitude and colors are averaged from \citet{co72,fe69,fe82,fe73,fe80,he76,tu80,va96}.\\
$^{a}$These stars were not used to plot the reddening map (Figure~\ref{rv_spt}) due to their known
variability or the uncertain intrinsic color relation [HD 93162, O2.5 If*/WN6 \citep{so14}].\\
\end{tabular}
\end{table*}

We first excluded variables reported in the literature 
\citep{fe63,fe82,fe73,fe80} to avoid misleading the reddening 
law. In addition, some stars show excess emissions in NIR 
bands because of dust grains heated by these stars. We 
statistically excluded such stars using 
a sigma-clipping method, where 2.5 times the standard deviation 
was adopted. The $E(V-\lambda)$ versus $E(B-V)$ diagrams 
for the early-type stars in the clusters are shown in Figure~\ref{rv_cl}. 
For Tr 14, Tr 16, and Cr 232, we obtained $E(B-V)_{fg}=0.35\pm0.02$ and $R_{V,cl}=4.40\pm0.27$ by fitting the 
reddening of the individual early-type stars to Equation 2. These results 
are almost the same values obtained in HSB12. On the other hand, 
we obtained a smaller value of $R_{V,cl}=3.43\pm0.11$ for Tr 15. 
If we do not consider the component of the foreground reddening, the slopes 
of the color excess ratios are consistent with that of the foreground 
component ($R_V=3.28\pm0.05$). This accords with the result of 
\citet{fe80} for Tr 15 ($R_V=3.2$).

There are several stars that were not 
observed in HSB12 and SSO2 because they are too bright or 
out of the FOVs. In order to investigate the reddening 
law out of the FOVs of the SSO data and 
determine the distance to the Carina Nebula including those stars,
we used the mean values of the photometric data 
from previous studies whose photometric data agreed well with ours 
\citep{co72,fe69,fe82,fe73,fe80,he76,tu80,va96}. 
The mean $V$ magnitudes, $U-B$, $B-V$, and $V-I$ colors are 
summarized in Table~\ref{star_sat}. 

We confirm that the foreground reddening law derived 
from the foreground early-type stars [$E(B-V)<0.35$] 
well follows the standard one ($R_{V,fg}=3.1$). Figure~\ref{rv_spt} displays the 
spatial variation of $R_{V,cl}$ and a uniform 
foreground reddening was assumed in the figure. Around 
Tr 14 and Tr 16, the mean $R_{V,cl}$ is $\sim4.5$ and in 
some regions $R_{V,cl}$ reaches up to 
$\sim4.8$, while the $R_{V,cl}$ around Tr 15 becomes to 
the normal value of $R_V=3.1$. A 
large $R_V$ means that the total extinction ($A_V$) 
overwhelms the reddening (or color excess values), 
indicating that the process of absorption by large dust 
grains effectively occurs compared to dust scattering. 
Therefore, the size of dust grains in the southern 
clusters Tr 14, Tr 16, and Cr 232 is still large.

\section{Distance\label{dist}}
The distance to the Carina Nebula has been a 
controversial issue, especially for the distance 
determination from the classical ZAMS fitting technique 
(see \citealt{wa95} and HSB12). The imperfect reddening 
correction is a major cause. HSB12 adopted the abnormal 
reddening law and obtained the photometric distance to 
the Carina Nebula to be $2.9 \pm 0.3$ kpc. However, 
\citet{LNGR19} and \citet{sh21} obtained the distance the 
Carina Nebula to be about 2.3 -- 2.4 kpc from Gaia parallaxes, 
which is in good agreement with the distance from parallax 
of the Humunclus Nebula \citep{sm06}. In this study, we 
derived the distance to clusters in the Carina Nebula 
using a reddening-free ZAMS fitting technique.

Most clusters show weak concentrations 
compared to Tr 14 as shown in the spatial distribution of 
early-type members (Figure~\ref{rv_spt}). In addition, there 
are a distributed stellar population across the nebula. We 
visually determined the areas of individual clusters given 
the spatial distribution of cluster members (see the circles 
in Figure~\ref{rv_spt}). The cluster areas properly contains 
most cluster members.

The reddening-free $Q^{\prime}$ magnitude and $Q$ colors that we 
adopted are expressed as below (see also \citealt{sos0}):
\begin{equation}
\begin{split}
Q' = (U-B)-0.72(B-V)-0.025E(B-V)^2 \\
Q_{VJ}=V-1.33(V-J) \\
Q_{VH}=V-1.17(V-H) \\
Q_{VK_S}=V-1.10(V-K_S)
\end{split}
\end{equation}

\begin{figure}[ht]
\includegraphics[width=1.05\columnwidth]{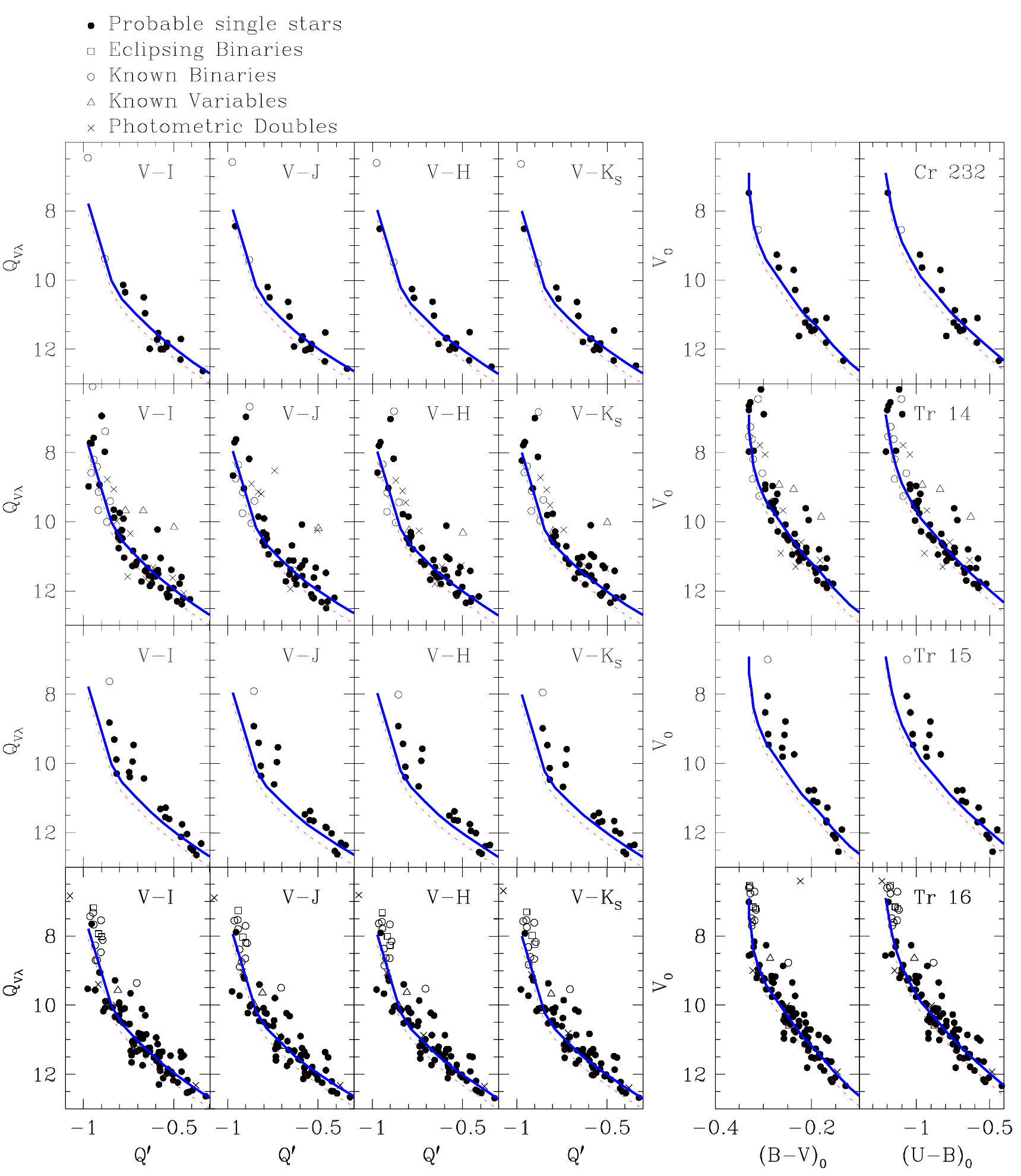}
\caption{
Reddening-free (left four panels) and reddening-corrected 
(right two panels) CMDs of the clusters in the Carina Nebula.
Open circles, triangles, and squares are confirmed or suspected 
binaries, variables, and eclipsing binaries from \citet[and references 
therein]{fe63,fe82,fe73,fe80,ra09, so14}, respectively. Photometric 
doubles and stars with $\epsilon_{(U-B)}>0.03$ or $\epsilon_{(B-V)}>0.03$ 
are marked with crosses. The blue thick solid and red thin dashed lines 
represent the ZAMS shifted by $V_0-M_V=11.9$ mag ($d=2.4$ kpc) 
and $V_0-M_V=12.2$ mag ($d=2.75$ kpc), respectively.}
\label{zams_fit}
\end{figure}

The $Q$ and $Q^{\prime}$ values of the early-type stars
were computed using he ZAMS relations from \citet{sos0}. 
We fit the ZAMS relations to the observed ($Q^{\prime}$, 
$Q_{V\lambda}$) diagram as shown in Figure~\ref{zams_fit}. Although 
clusters in the Carina Nebula are very young ($\lesssim3$ Myr for Tr 14 
and Tr 16, see HSB12), some early-O stars appear brighter than the 
magnitude of the ZAMS at a given color due to their rapid evolution.
In addition, the binary frequency of O-type stars is known to be 
very high \citep{sa12,so14}. If we fit the 
ZAMS relation to the bright part of the $Q^{\prime}$, $Q_{V\lambda}$
diagram, then the distance to individual clusters can be underestimated. 
Therefore, the ZAMS relation should be fit to the lower-ridge line 
of the faint and dense main sequence band at given 
colors following the original definition of the ZAMS \citep{sos0}.

Since the slope of the ZAMS relation for early-type stars is very 
steep (large changes in magnitude for small changes in color), 
even small photometric errors on color result in large errors on 
distance modulus. We only used early-type stars with very small 
photometric errors [$\epsilon_{(B-V)}\leq0.03$ and $\epsilon_{(U-B)}\leq0.03$] 
in the ZAMS fitting. The best-fit distance modulus is $V_0-M_V=11.9$ 
mag ($d=2.4$ kpc). We also display the results of ZAMS fitting in the 
reddening-corrected CMDs in the right panels of Figure~\ref{zams_fit} 
for comparison. However, there are some stars fainter than the 
ZAMS lines at given $Q$ colors. 
It is not easy to measure the error of the ZAMS fitting.
We note that an error of $0.03$ in $U-B$ can result in the errors of $Q^{\prime}$ magnitude 
and $(U-B)_0$ color, ultimately leading to a higher distance modulus (farther distance) up to $\pm0.3$ mag ($\pm 0.35$ kpc). 

We did not fit the ZAMS relation to the bright stars ($V_0 < 10$ mag) in Tr 15 because they may not be zero-age MS stars. Tr 15 is the oldest cluster ($\sim$6 Myr, \citealt{fe80}) and the brightest star in the cluster (HD 93249, O9 III) is a definitely evolved from MS, while earlier MS stars (O3.5 V stars) are still exist in Tr 14 and Tr 16. But the fainter ($V_0 > 10$ mag) stars fit well to the ZAMS relation. In addition, a recent study \citep{sh21} also supports our result that Tr 15 is at the same distance as Tr 14 and Tr 16.

\begin{figure}[ht]
\includegraphics[width=\columnwidth]{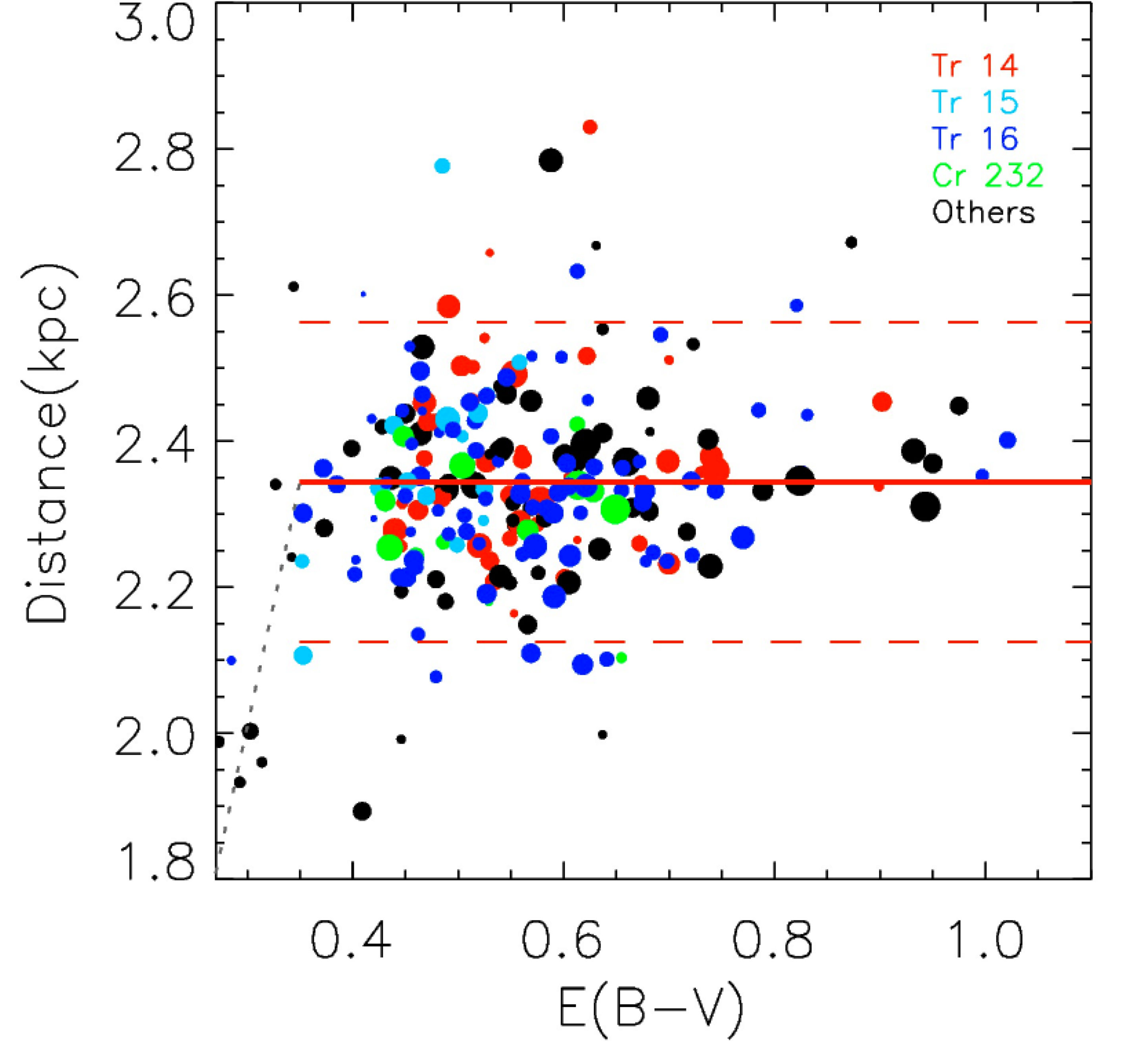}
\caption{Distance distribution of the early-type stars. The 
distances were obtained from the inversion of the Gaia parallaxes \citep{edr3}. The larger dots indicates small errors in both 
photometry and parallax. The red lines 
indicate the mean distance ($2.34$ kpc, solid) and the 
2.5$\sigma$ limits (long dashed). The gray line (short dashed) 
shows the foreground reddening ($E(B-V)=0.15$ mag/kpc).}
\label{dist_gaia}
\end{figure}

The distance to the Carina Nebula was also derived from the inversion 
of the Gaia parallaxes. Figure~\ref{dist_gaia} shows the distance 
distribution of the early-type stars with respect to their reddening 
values. For the foreground stars ($E(B-V)<0.35$), reddening linearly increases 
by approximately $E(B-V)\sim0.15$ mag/kpc, while there is no clear 
trend but with a large scatter for the Carina Nebula ($E(B-V)\geq0.35$).
We took the mean distances after eliminating the some outliers using 
a sigma-clipping method, where parallaxes smaller than 2.5 times the standard 
deviation were adopted. The distance to the Carina Nebula from the Gaia parallaxes 
is $2.34\pm0.09$ kpc from 214 stars (193 calculated, 21 rejected), which 
is consistent with the distance derived from the ZAMS fitting method. 
Our results are in good agreement with the results of some previous 
studies ($\sim$ 2.3 kpc) within errors \citep{al93,me99,da01,sm06,re12,LNGR19,
sh21}. 

\begin{figure}
\includegraphics[width=\columnwidth]{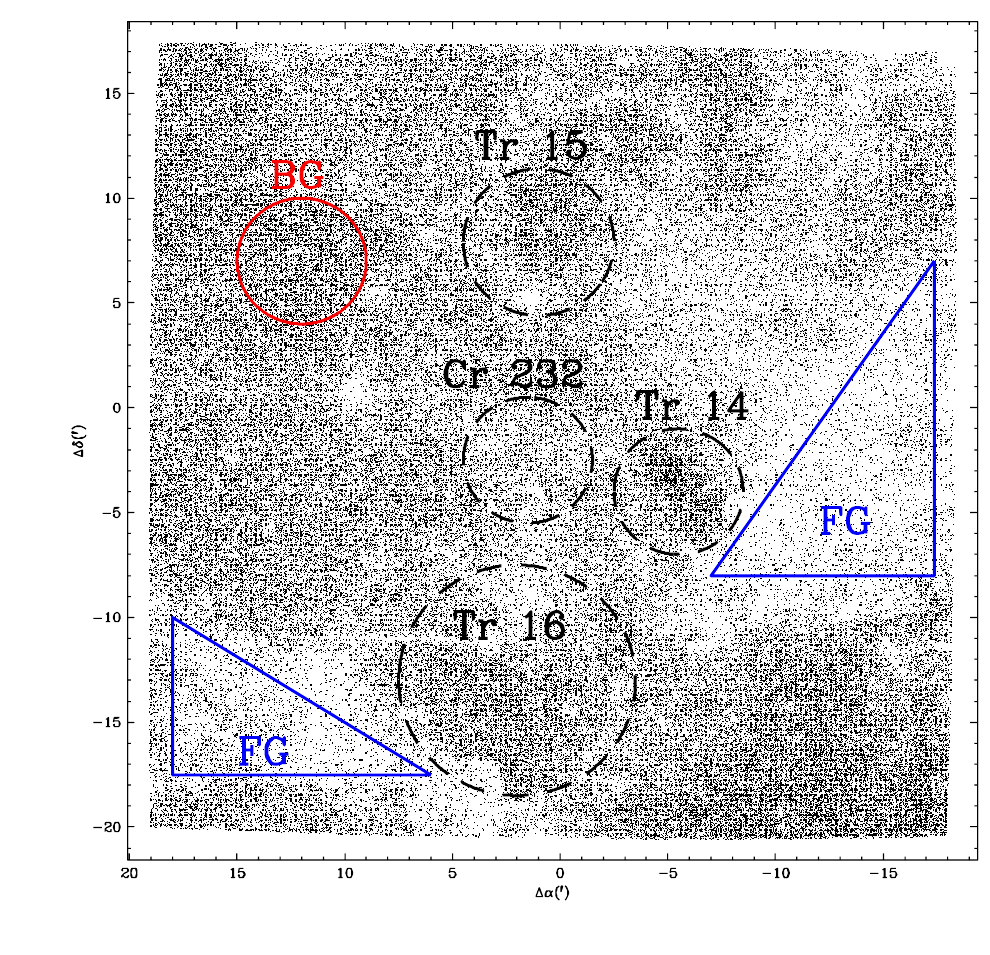}
\caption{
The finder chart for all the detected stars. 
The circular regions outlined by dashed and solid lines 
represent the adopted boundary of the four young clusters 
and the background (BG) field, respectively. The FG fields are shown by 
triangles.}
\label{chart_region}
\end{figure}

\begin{figure*}
\includegraphics[angle=270,width=2.0\columnwidth]{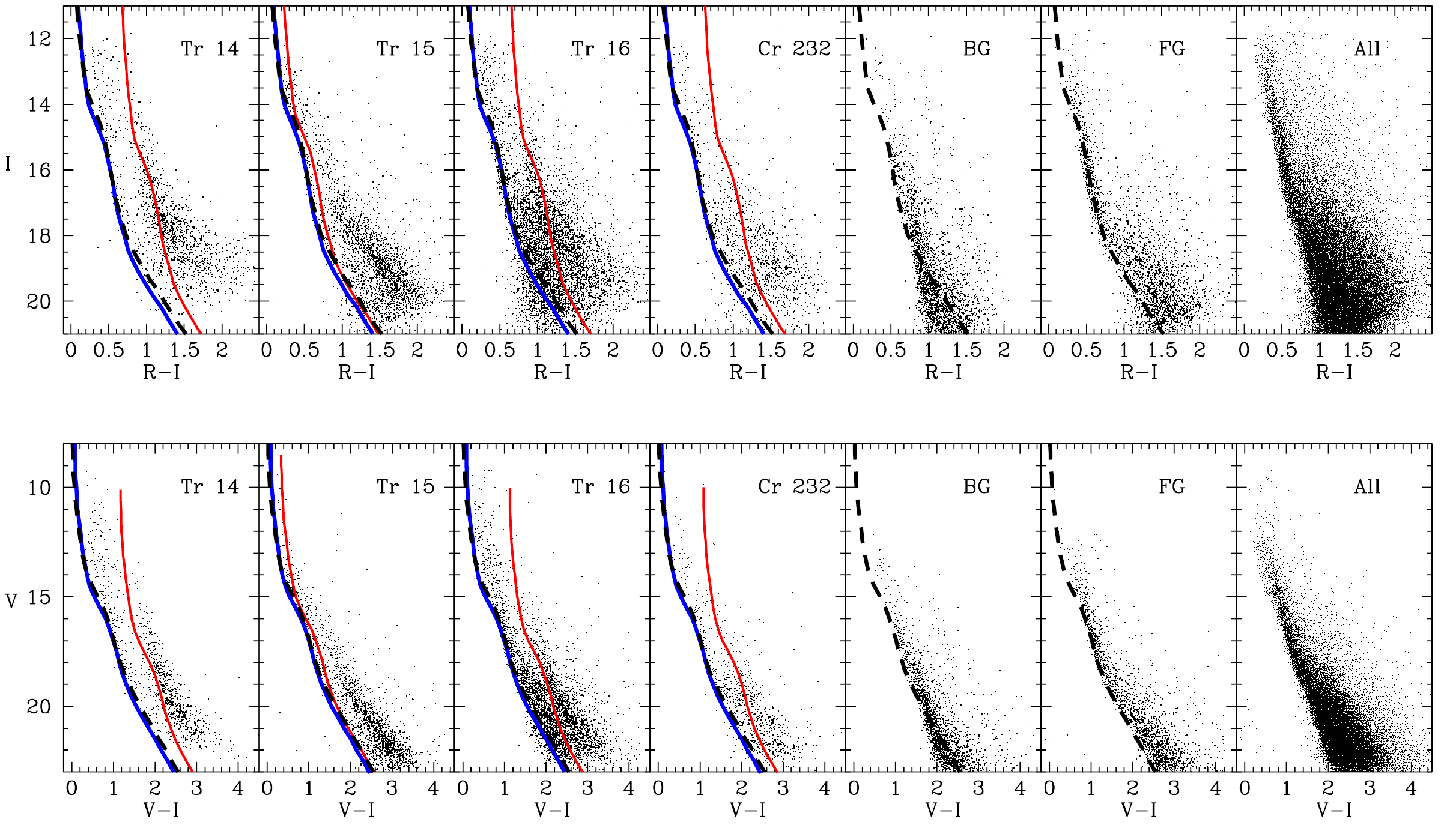}
\caption{
Color-magnitude diagrams of the clusters and selected fields. The ZAMS lines are shown for comparison to PMS, FG, and BG stars.
The black curves shown by dashed lines is the ZAMS relations shifted 
by $E(B-V)=0.3$ and $V_0-M_V=11.5$ mag ($d=2.0$ kpc) to roughly 
fit the foreground stars. The blue curves 
represent the ZAMS relations reddened by the lowest reddening 
value of $E(B-V)=0.35$ (all panels), while the red curves are the 
ones reddened by the high reddening values of $E(B-V)=0.8$ 
(0.5 for Tr 15) and the $R_{V,cl}$ values of given clusters. The same distance 
mudulus of 11.9 mag was applied to these ZAMS relations (blue 
and red curves).}
\label{cmd_all}
\end{figure*}

\section{Photometric Diagrams\label{pd}}

Figure~\ref{chart_region} displays the 
spatial distribution of all detected sources in the FOV of 
our data. The surface number density of stars varies spatially, 
which indicates the 
inhomogeneous distribution of interstellar matter toward 
the Carina Nebula. We assigned two field showing the lowest 
surface density to foreground fields (FG). The FG fields 
cover the V-shaped dark clouds (see also 
Figure~\ref{chart}), which seem to block the stellar light 
from the members of the Carina Nebula as well as that from 
background stars. On the other hand, one field where the 
surface density of faint stars is the highest outside the 
clusters was selected as the BG field. 

We present the CMDs of the clusters, the BG field, the FG 
fields, and all stars in Figure~\ref{cmd_all}. It also 
supports that the different completeness levels from region 
to region are also partly related to the inhomogeneity of 
interstellar matter. FG contamination in the Carina Nebula 
is expected to be uniform \citep{ge11}, while the BG field 
is dominated by FG and BG stellar populations. Hence, it 
is necessary to probe the photometric properties of stellar 
populations in the three fields to better select genuine 
members from the observed photometric diagrams.

In the CMDs, while the maximum foreground 
reddening $E(B-V) = 0.35$ and the distance modulus of the 
clusters were adopted for the less reddened ZAMS relation 
(blue solid curves), the more reddened ZAMS (red solid curves) 
is shifted by mean reddening values and their distance moduli. 
We roughly fit the ZAMS relation to the foreground population 
(black dashed curves) with [$E(B-V)=0.3$, $V_0 - M_V = 11.5$]. 
In the CMDs for the BG field, two distinct sequences are seen, 
one is the most populous sequence in the CMDs, while the other 
is a weak, but clear red sequence from ($V-I, V$) $\sim$ ($2.0,15$) 
and ($R-I, I$) $\sim$ ($1.0,13$) to ($V-I, V$) $\sim$ 
($4.0,23$) and ($R-I, I$) $\sim$ ($1.8,21$). The former population 
seems to be the field MS star population because the number 
of stars implies the lifetime of stars. Meanwhile, 
the latter population seems to be either red giants or 
red clump stars, given some factors, such as the number of these 
stars smaller than those of the MS stars, the reddening vector, and 
the effects of different distances along the line of sight.

\begin{figure*}[ht]
\includegraphics[angle=90,width=2.05\columnwidth]{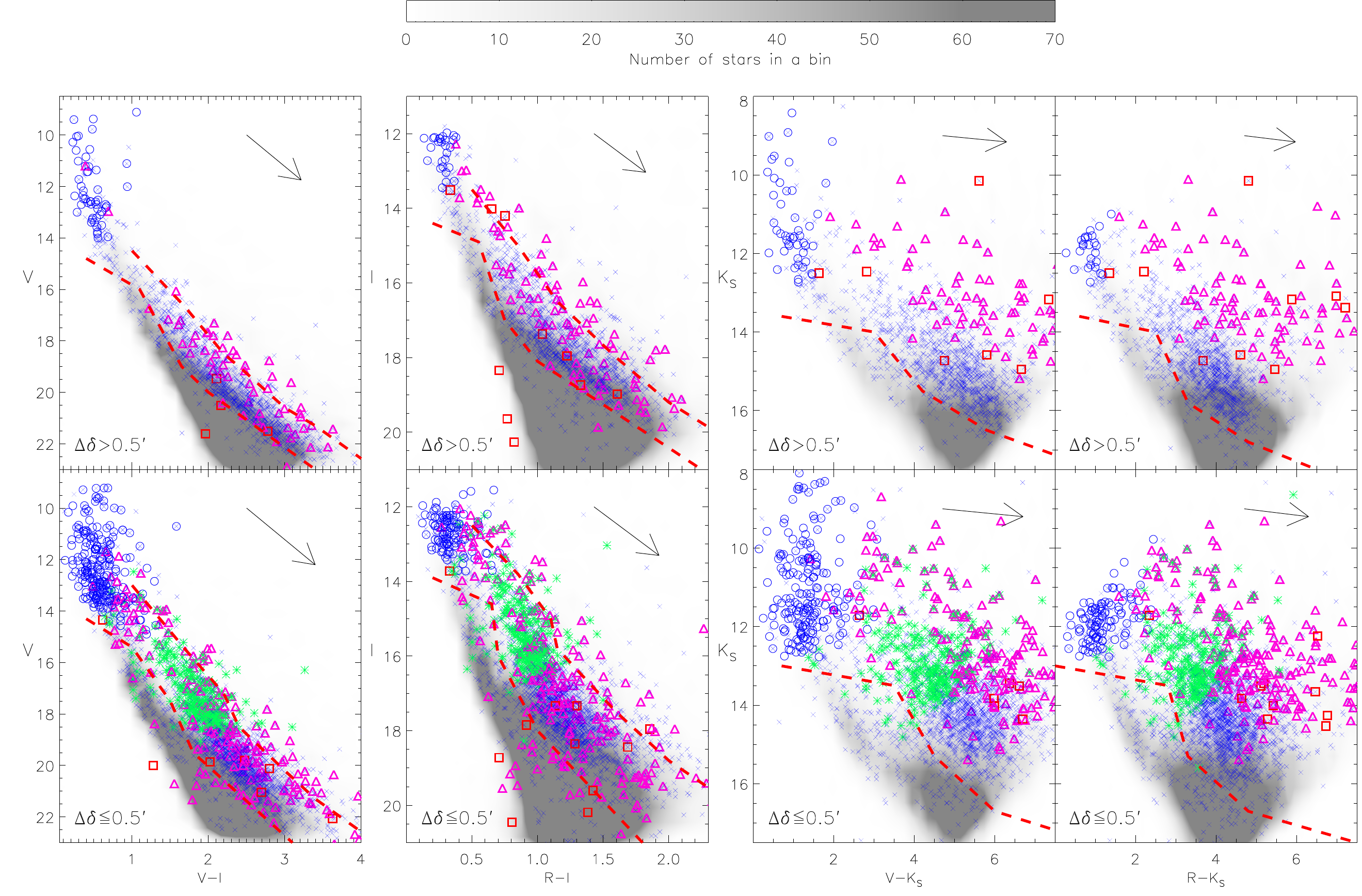}
\caption{
Color-magnitude diagrams of stars in the southern ($\Delta\delta\leq0.5$ arcmin, bottom panels) 
and northern ($\Delta\delta >0.5$ arcmin, upper panels) regions. The symbols of
stars with the membership criteria are the same as in Figure~\ref{ubbv}. The field
interlopers are shown as the grey-scale density map. The thick dashed curves in the
($V,V-I$) and ($I,R-I$) CMDs are the PMS loci, while the thick dashed curves
in the ($K_S,V-K_S$) and ($K_S,R-K_S$) CMDs are the lower boundaries of
members in the Carina Nebula. The arrows indicate the reddening vectors
of $E(B-V)=0.5$ for $R_{V,cl}=4.5$ (bottom panels) and $R_{V,cl}=3.5$ (upper panels).}
\label{cmd_m}
\end{figure*}

In Tr 14 and Tr 15, there are the prominent sequences of 
numerous fainter ($V>16$ mag and $I>16$ mag) and redder 
($V-I > 2$ and $R-I > 1$) stars. A large fraction of these 
stars may be PMS stars in the clusters because many of 
them are X-ray emission stars (see also Figure~\ref{cmd_m}). 
On the other hand, the CMDs of Tr 16 show a complex morphology. 
The redder sequence is very similar to the PMS sequence of Tr 14, 
but the bluer sequence is similar to that of background stars 
as seen in the CMDs of the BG field. The 
morphology of Cr 232 CMDs are very similar to that of Tr 14,
although the number of stars are much smaller. This result 
supports again that Cr 232 is an extend part of Tr 14 
\citep{ta03,ca04};HSB12.

\subsection{PMS Member Distribution in Color-Magnitude Diagrams\label{cmd_pms}}

Figure~\ref{cmd_m} shows the optical (left) and optical-NIR 
(right) CMDs with the loci of probable members. Although the 
stars in the Carina Nebula are relatively concentrated in/around 
three major clusters (Tr 14, Tr 15, and Tr 16), many stars in the 
loci of members are found outside the clusters and over the whole 
nebula region. The spatial distribution of 
members around the clusters is very complex, and therefore it 
is very difficult to define the boundaries of clusters. \citet{fe11} 
also claimed that the stellar clusters in the Carina Nebula are 
difficult to separate from each other by simple division. In 
addition, HSB12 argued that Cr 232 seems not a independent 
cluster from Tr 14 and Tr 16. We simply divided 
the observed FOV into two regions, the southern 
($\Delta\delta\leq 0.5$) and northern ($\Delta\delta >0.5$) 
regions, instead of individual clusters, in order to check 
the global distribution of X-ray emission stars and MIR excess 
stars in the CMDs. This division is based on the mean 
$R_{V,cl}$ values of $\sim4.5$ (southern) and $\sim3.5$ (northern). 
The southern region contains Tr 14, Tr 16, and Cr 232, while the 
northern region contains Tr 15 that is known to be the oldest 
cluster (several Myr, \citealt{sb08}).  

It is necessary to compare the completeness limit of our data 
with that of the CCCP survey because our member selection is 
highly relied on its X-ray source list. The number of stars 
with X-ray counterparts in the less reddened northern region 
clearly decreases in the low-luminosity regime of the CMDs 
($V\gtrsim 22$ mag and $I\gtrsim 19.5$ mag). The detection 
limit of the CCCP sources in our data is calculated to be 
about $0.4$ M$_{\odot}$ in $V$ and $0.2$ M$_{\odot}$ in $I$ 
[$\log (L/L_{\odot})= 0.06$] adopting the cluster age of 7 Myr, 
$R_{V,cl}=3.5$, and $E(B-V)= 0.6$. \citet{br11b} found that 
the X-ray sources in the CCCP catalog is complete down to 
$\log L_X=30.7$ (erg sec$^{-1}$) at $d=2.3$ kpc. If we 
assume the saturation level of X-ray luminosity due to the 
dynamo action ($\log (L_X/L_{Bol}) = -3$), the limiting 
luminosity estimated from the CCCP survey is then about 
$\log (L/L_{\odot})\sim 0.1$, which is consistent with the 
completeness limit of our optical data. Therefore, our data 
fully cover the low-mass limit of the CCCP survey for the 
northern region.

Figure~\ref{cmd_m} exhibits the CMDs of stars. Many X-ray 
emission stars in the ranges of $V-I= 0.4$--$1.0$, 
$R-I= 0.2$--$0.5$, and $V<14.5$ mag may be early-type 
MS members in the Carian Nebula. A large number of PMS 
stars with X-ray emission or MIR excess emission constitute 
a sequence in ($V, V-I$) and ($I, R-I$) CMDs (thick dashed 
lines in the left panels of Figure~\ref{cmd_m}). We have 
referred to this sequence as PMS loci. This is very useful 
to search for additional member candidates without neither 
X-ray emission nor MIR excess emission. In addition, there 
are several class 0/I or II stars below the PMS loci in the 
($V, V-I$) and ($I, R-I$) CMDs, while there are no class 
0/I or II stars below the lower boundary in the ($K_S, V-K_S$) 
and ($K_S, R-K_S$) CMDs. These stars may be classical T Tauri 
stars with thick edge-on circumstellar disks \citep{su09}.

We also found a number of X-ray emission stars below the lower 
boundary of the PMS loci ($V-I= 0.8$--$1.5$ and $V= 16$--$19$ 
mag, or $R-I= 0.3$--$0.7$ and $I= 14.5$--$18$ mag). If the 
foreground reddening of $E(B-V)=0.3$ and $V_0-M_V=11.5$ 
mag (HSB12) are assumed, the ZAMS relation is well fit to the 
CMDs of such stars. They may be foreground FG-type MS stars 
rather than PMS members. There are 19 LH stars below the lower 
boundary of the PMS loci. Unlike class 0/I/II stars, these 
LH stars are below the PMS loci in both optical and optical-NIR 
composite CMDs. About half (8/19, 42\%) of these stars show a 
weak (EW $< 150$ m\AA) or no absorption of Li I $\lambda$6708. 
Therefore, the LH stars below the PMS loci may be young stars 
in the foreground, but older than the YSOs in the Carina Nebula.

In the CMDs, the number density of field interlopers is very 
high near the lower boundary of the PMS loci. These stars are 
major contamination sources in selection of PMS members. Some of 
foreground stars can be rejected from the ($K_S,V-K_S$) and 
($K_S,R-K_S$) CMDs because more reddened stars are clearly 
separated from less reddened, intrinsically red stars in the 
optical-NIR composite CMDs than in the optical CMDs due to 
the differences between the reddening vectors of the optical 
and NIR wavelengths. Since some PMS stars with active accretion 
disks may appear bright in the $K_S$ band, we only set the lower 
boundary in the ($K_S,V-K_S$) or ($K_S,R-K_S$) CMD to exclude 
foreground interlopers in the selected PMS candidates.

\begin{figure}[ht]
\includegraphics[width=\columnwidth]{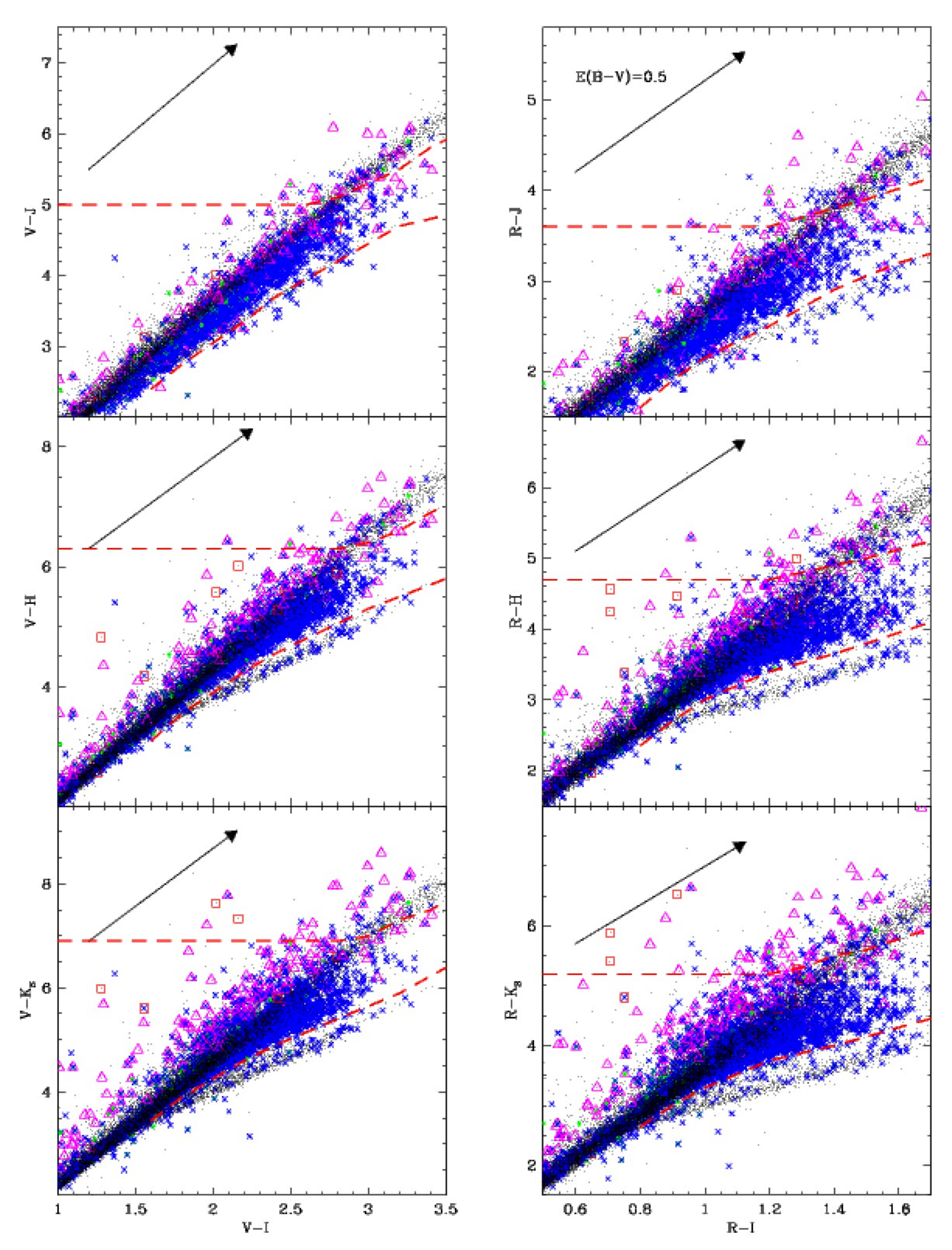}
\caption{Optical-NIR versus optical TCDs.The symbols are the same as those in Figure~\ref{cmd_m}. The lower and upper thick dashed lines (red) are the borderlines for the exclusion of foreground dwarfs and background giants, respectively.
The arrows indicate the direction of reddening vector in each diagram for $E(B-V)=0.5$ mag and $R_{V,cl}=4.5$.}
\label{ccd_vrijhk}
\end{figure}

\subsection{Two-Color Diagrams}
 We used the optical-NIR TCDs to probe the distribution 
of the field interlopers (Figure~\ref{ccd_vrijhk}). We divided the diagrams into three sections.
Most of the X-ray emission stars occupy the mid-section.
The upper (redder in the optical-NIR colors) section includes the stars extended
along the reddening vector, and therefore seems to be highly reddened 
background stars. The Carina Nebula members with IR-excess may also 
be in the upper section. The bluer sequence 
bifurcated from $V-I \sim 1.6$ and $R-I \sim 0.8$ (foreground MS stars) is clearly separated from the member stars by the lower border line shown in Figure~\ref{ccd_vrijhk} (see also figures 5 and 12 in \citealt{wl11} 
for the intrinsic color-color relations of MS and giant). 

\begin{figure}[ht]
\centering
\includegraphics[width=0.7\columnwidth]{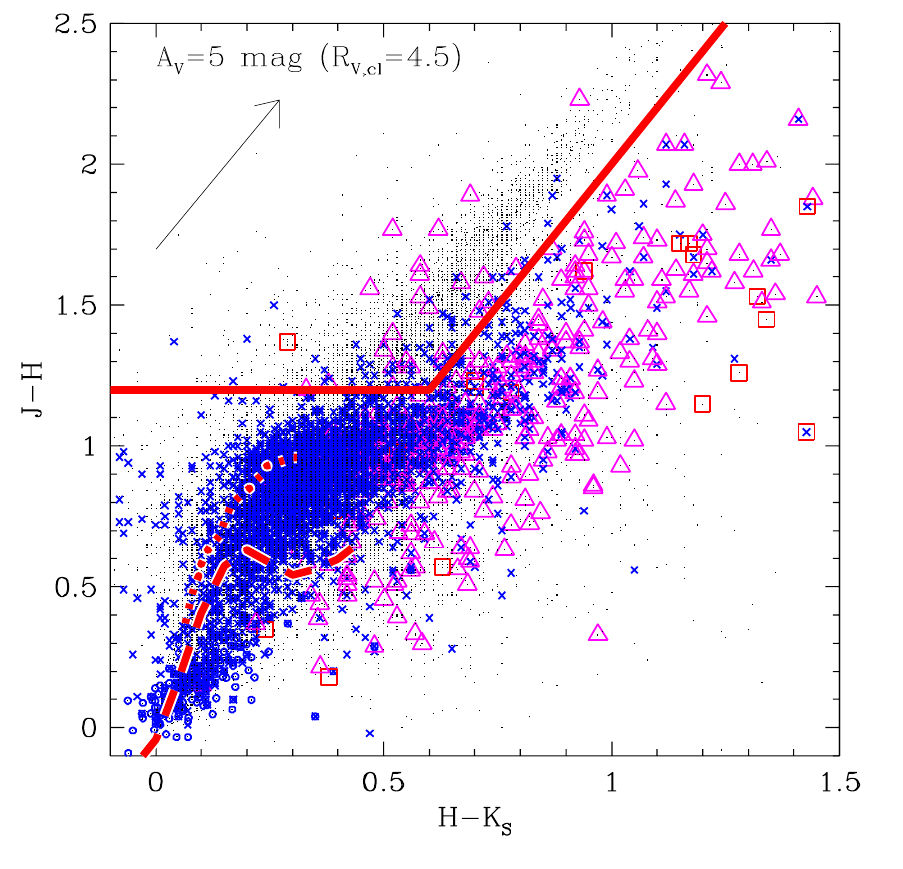}
\caption{
($J-H,H-K_S$) TCD for the stars detected in the optical bands with $\epsilon[\equiv\sqrt{\epsilon_{(J-H)}^2+\epsilon_{(H-K_S)}^2} ]\leq 0.1$.
The dots (black) are stars without any membership criterion, and the other symbols are the same as in Figure~\ref{cmd_m}.
The long and short dashed lines are the MS relation from \citet{su08} and the intrinsic color-color relation for giants from \citet{be88}.
The thick solid line represents the borderline for the exclusion of reddened giants in the membership selection.}
\label{ccd_jhk}
\end{figure}

Asymptotic giant branch (AGB) stars with MIR excess can 
also be found in the same loci as PMS stars in the 
optical-NIR CMDs. AGB star candidates can be identified in 
the ($J-H,H-K_S$) TCD (Figure~\ref{ccd_jhk}). There is 
a sequence of giant stars along the reddening vector in the 
color range of ($H-K_S, J-H$)$\sim$(0.5, 1.2) to (1.0, 2.2). 
We set the borderline (thick solid line) to identify highly 
reddened giants including AGB stars.

\subsection{PMS Membership Selection}
In the previous sections, we discussed 
stellar population and possible field interlopers toward 
the Carina Nebula. A number of stars found in the PMS loci 
are probable PMS members, especially for the X-ray sources 
and LH stars. The CCCP ``H1''--``H3'' sources and LH stars 
found within the PMS loci in the optical CMDs as PMS members. All stars with MIR excess emission 
were also included in the PMS member list.
 
There are several X-ray emission stars brighter than the 
PMS loci. These stars may be either X-ray active stars in 
the foreground or PMS members whose magnitude is brighter 
than the average luminosity of normal PMS stars. For instance, 
actively accreting PMS stars may be brighter than normal 
PMS stars if their accretion luminosity is as bright as 
their stellar luminosity ($L_{\text{acc}} \sim L_*$, 
\citealt{na06,li14}). The binary fraction of PMS stars 
in dense young open clusters has not been investigated well, 
but the PMS stars of the young open cluster NGC 2362 
($\sim$5 Myr with a very small intracluster reddening) shows 
a clear equal-mass binary sequence \citep{mo01,da05}. 
In addition, as many stars in young open clusters show 
a spread in age of a few Myr \citep{su10,li16}, we can 
expect to find some bright PMS stars younger than the 
rest. Therefore, several X-ray emission stars and LH 
stars were also included in the PMS members even if 
they are brighter by up to one magnitude than the upper 
boundary of the PMS loci.

The other stars without any signature of young stars 
found in the PMS loci were classified as PMS candidates.
A total of 3,594 PMS members and 23,147 PMS candidates 
were selected from the CMDs. However, a number of 
nonmembers may be included in the lists of members and 
candidates. We evaluated the membership of the selected 
PMS stars and candidates based on the distributions 
of field interlopers in CMDs and TCDs as discussed 
in previous sections. From the NIR-optical CMDs, a total of 
129 and 3,801 stars are excluded from PMS members and candidates, respectively, as they are located below the lower boundary in the 
($K_S, V-K_S$) and ($K_S, R-K_S$) CMDs. From the 
Optical-NIR TCDs, stars above the upper boarder lines without IR-excess and all stars below the lower boarder lines 
are excluded from the members and candidates. A total of 68 and two PMS members are re-classified as foreground and background stars, respectively. In addition, 
a total of 403 and 3,806 PMS candidates were re-classified as foreground and background stars, respectively. 
In the ($J-H,H-K_S$) TCD, we found that 64 PMS members and 163 PMS candidates follow the sequence of highly reddened 
giants. These stars were left out of the member list. 
Finally, our member list contains 3,331 PMS members and 
14,974 PMS candidates.  

\begin{figure*}[ht]
\includegraphics[width=1.2\columnwidth]{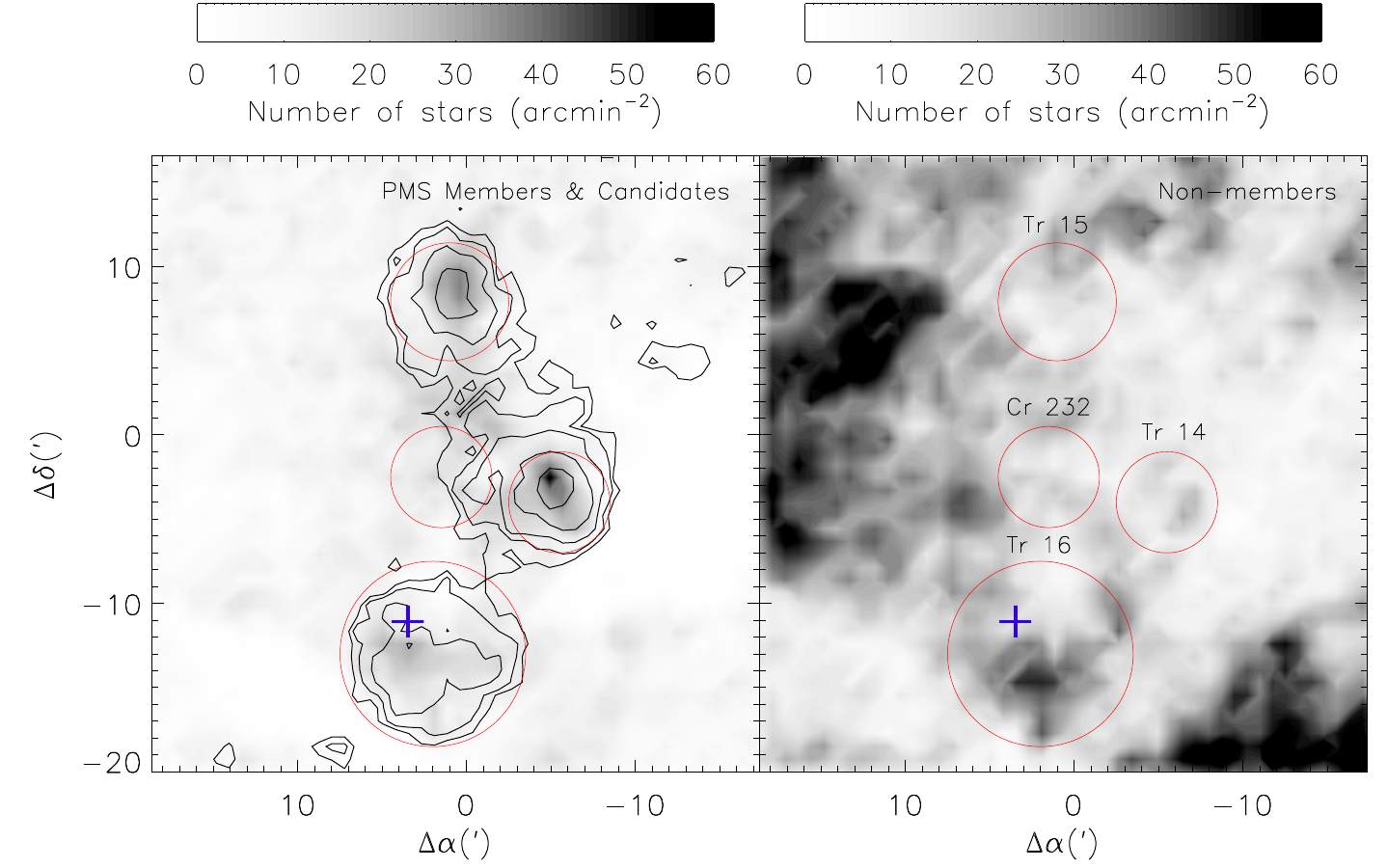}
\includegraphics[width=0.87\columnwidth]{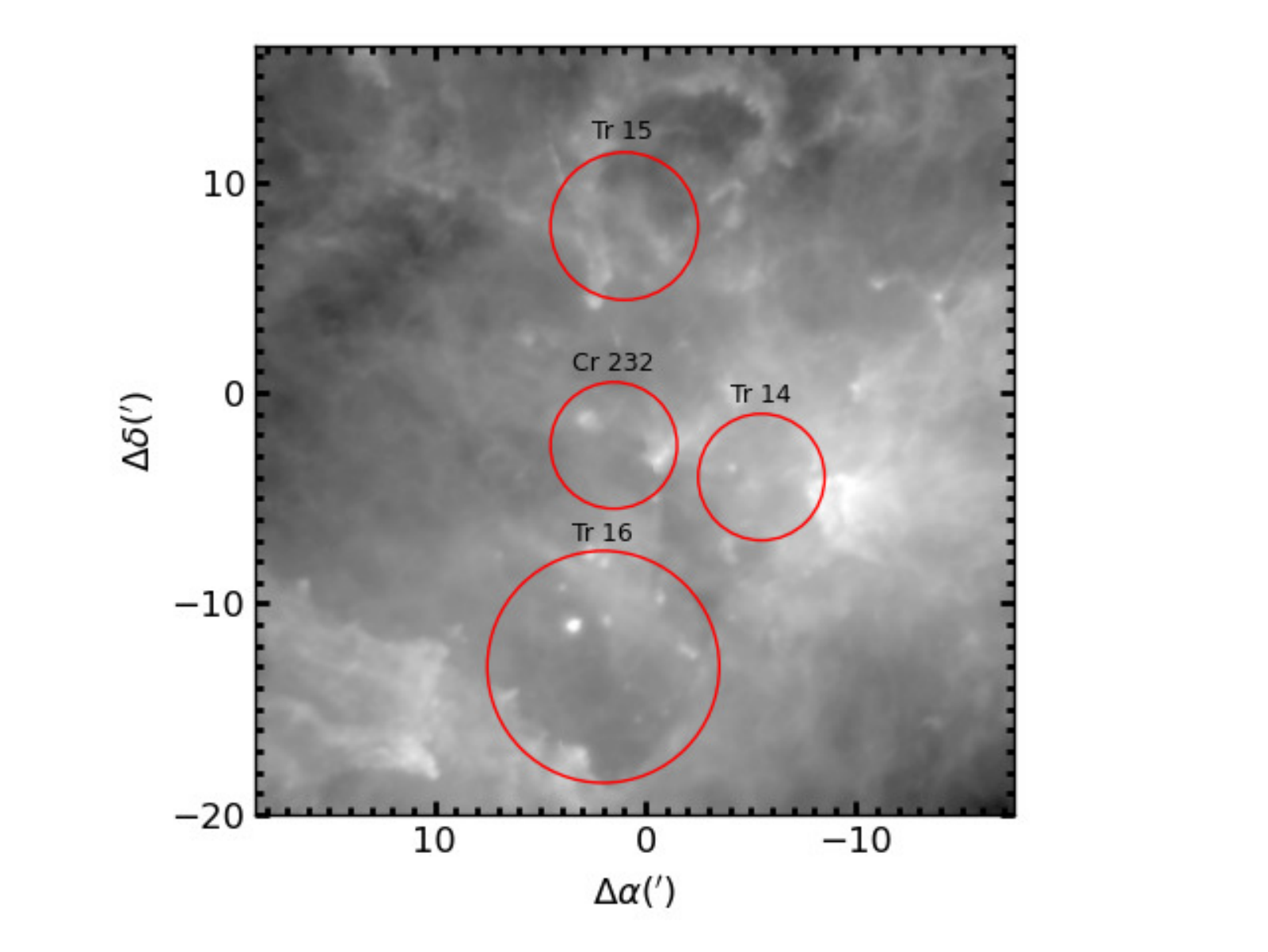}
\caption{Surface density map of the members and candidates (left), 
non-members (middle), and Herschel PACS 160$\mu$m image 
(right; \citealt{PRP10,PWG10}). 
The contours in the left panel indicate the surface densities of 3, 5, 10, 20, and 40 members per arcmin$^2$. The gray scale in the left panel traces the surface densities of PMS candidates, while middle panel shows the surface densities of non-members.
The red circles show the apparent locations and sizes of the major clusters and the cross denotes the position of $\eta$ Carinae. In the right panel, the Herschel PACS image directly shows the inhomogeneous 
distribution of warm dust toward the Carina Nebula.}
\label{spt_map}
\end{figure*}

\section{Spatial Distribution of the PMS Members and Candidates}\label{sec_spt}
Figure~\ref{spt_map} compares the spatial distributions 
of the PMS members (contour) with those of PMS candidates 
(gray scales) and nonmembers. The spatial distribution of 
PMS candidates are very similar to that of PMS members 
although the candidates have a wider distribution. It supports 
that our member selection is reliable. Interestingly, many 
PMS candidates are found between Tr 14 
and Tr 15. Their surface density in the region is not very 
high, however it is certain that there are a number of 
young stars, e.g., early-type stars, X-ray sources, and 
several jets \citep{sm10,ha15}. These candidates may be 
the halo stars of Tr 14. This structural feature is reminiscent 
of the northern clump of Westerlund 2 \citep{hur15}, the halos 
of IC 1805 \citep{LHY20}, NGC 2244 \citep{LNH21}, and NGC 
2264 \citep{LNH22}.

The spatial distribution of nonmembers is far different 
from that of PMS members and PMS candidates. The distribution of 
their surface density traces the inhomogeneous distribution 
of interstellar matter toward the Carina Nebula. The 
enhancement of surface density can be seen to the east 
and southwest of the Carina Nebula (see also 
Figure~\ref{chart}). Herschel\footnote{Herschel is an ESA space observatory with science instruments provided by European-led Principal Investigator consortia and with important participation from NASA.} Photodetector Array Camera 
and Spectrometer (PACS -- \citealt{PRP10,PWG10}) image at 160$\mu$m shows the 
distribution of warm dust (bright regions in the right 
panel of Figure~\ref{spt_map}). The east and southwest of 
the Carina Nebula appear to be darker than the other 
regions in the Herschel image, which means that these 
two regions are transparent. It confirms that a large number 
of stars in the eastern region are background stars.

\begin{figure}
\centering
\includegraphics[width=0.7\columnwidth]{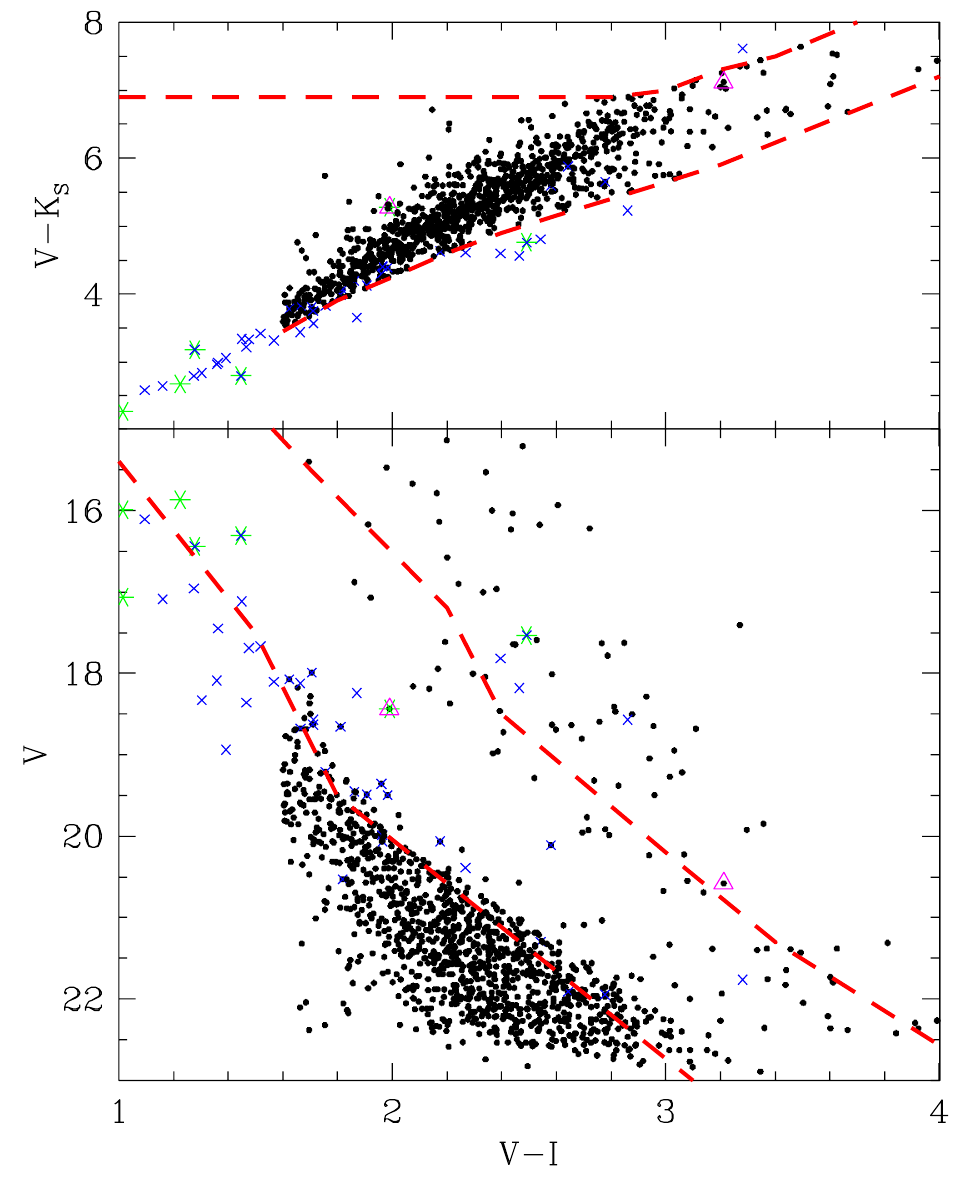}
\caption{
The CMD (lower panel) and TCD (upper panel) of non-members in Tr 16.
The locus of PMS members is indicated by the dashed lines.
The stars with $V-I>1.6$ and within the upper and lower limits in the TCD are highlighted as large dots.
The other symbols are the same as in Figure~\ref{ccd_jhk}.
\label{Tr 16_nonmem}}
\end{figure}

A small level of the density enhancement of nonmembers 
are found toward Tr 16. The CMD and TCD of the stars classified 
as nonmembers in Tr 16 are shown in Figure~\ref{Tr 16_nonmem}.
In the figure, most of those nonmembers are distributed within the 
the boundaries outlined by dashed curves in the ($V-K_S$, $V-I$) 
TCD, and they are fainter than the lower boundary of the PMS locus 
in the ($V$, $V-I$) CMD. Therefore, a large fraction of nonmembers 
toward Tr 16 may be background stars. Indeed, the position of 
the density enhancement corresponds to the cavity seen in far-infrared, sub-mm, and CO intensity maps \citep{yo05,pr11b,pr12} and the right 
panel of Figure~\ref{spt_map}. 

\section{Summary and Conclusion}
We presented a deep homogeneous $UBVRI$ photometry 
of stellar population toward the Carina Nebula. In order to minimize 
false detections, all images were carefully checked by visual 
inspection. The 90\% completeness limit near Tr 15, where the 
nebula is faint, is $V=22.4$, $R=21.8$, and $I=21.1$ mag, while
the completeness limit becomes much brighter in the brightest 
part of the Carina Nebula, down to $V=20.7$ mag and $I=20.3$ mag near Tr 14 and $\eta$ Carinae.

We found that $R_{V,cl}$ of the intracluster medium systemically 
decreases from south (Tr 14 and Tr 16, the central part of the Carina 
Nebula) to north (Tr 15). Our result indicates that the sizes of dust 
grains are different region by region. We revised the distance modulus 
of Tr 14 and Tr 16 to $V_0-M_V=11.9 \pm 0.3$ mag ($d=2.4 \pm 0.35$ kpc)
based on the updated ZAMS relation and $E(U-B)/E(B-V)$ ratio. Our 
photometric distance is in good agreement with the distance derived from 
the Gaia EDR3 parallaxes ($2.34\pm0.09$).

PMS members with X-ray emission or MIR excess were 
selected from cross-matches with the {\it Chandra} X-ray 
point source catalog from the CCCP and {\it Spitzer} MIR 
data from the Vela-Carina survey. 
The distribution of the X-ray emitting stars in the CMDs shows that
our optical data reach fainter than the CCCP detection limit.
We classified the PMS members and candidates and showed that
the PMS membership can be improved using multi-wavelength data 
combined with our deep optical data.

Our deep and homogeneous optical photometric data will be widely 
used by stellar and Galactic science communities. Furthermore, 
the list of members is very useful to study the properties of 
young stellar population, the star formation history in the 
Carina Nebula, and the initial mass functions of the stellar 
clusters (Hur et al. in prep.). The data presented here would 
have a legacy value for follow-up studies.

\acknowledgments
The authors thank the anonymous referee for many constructive 
comments and suggestions. The authors also wish to thank Professor 
Michael S Bessell for many comments and discussions. We dedicate 
this work to our colleague Hwankyung Sung. This paper has made 
use of data obtained from the European 
Space Agency (ESA) mission {\it Gaia} (https://www.cosmos.esa.int/gaia), 
processed by the {\it Gaia} Data Processing and Analysis Consortium 
(DPAC, https://www.cosmos.esa.int/web/gaia/dpac/consortium). Funding 
for the DPAC has been provided by national institutions, in particular the institutions participating in the {\it Gaia} Multilateral Agreement. PACS has been developed by a consortium of institutes led by MPE (Germany) and including UVIE (Austria); KU Leuven, CSL, IMEC (Belgium); CEA, LAM (France); MPIA (Germany); INAF-IFSI/OAA/OAP/OAT, LENS, SISSA (Italy); IAC (Spain). This development has been supported by the funding agencies BMVIT (Austria), ESA-PRODEX (Belgium), CEA/CNES (France), DLR (Germany), ASI/INAF (Italy), and CICYT/MCYT (Spain). This research has also made use of the SIMBAD database \citep{simbad}, operated at CDS, Strasbourg, France. 
This work was supported by the National Research Foundation of Korea (NRF) 
grant funded by the Korean government (MSIT) (Grant No : 2022R1C1C2004102).

\end{document}